\documentclass[10pt,journal,compsoc]{IEEEtran}
 
\ifCLASSOPTIONcompsoc
  \usepackage[nocompress]{cite}
\else
  \usepackage{cite}
\fi
\usepackage{amsmath,amssymb,amsfonts}
\usepackage{algorithmic}
\usepackage{graphicx}
\usepackage{textcomp}
\usepackage[hyphens]{url}
\usepackage{balance}
\usepackage{mathptmx}
\usepackage{fancyhdr}
\usepackage{makecell}
\usepackage[normalem]{ulem}
\usepackage[hyphens]{url}
\usepackage[keeplastbox]{flushend}
\usepackage[bookmarks=true,breaklinks=true,colorlinks,linkcolor=black,citecolor=blue,urlcolor=black]{hyperref}
\usepackage{float}
\usepackage{makecell}
\usepackage{comment}  
\usepackage[table]{xcolor}
\usepackage{pgf}
\usepackage{collcell}
\usepackage{booktabs}
\usepackage{multirow}
\usepackage{pifont}%
\usepackage[ruled,linesnumbered]{algorithm2e}
\usepackage{soul}
\usepackage{pifont}
\usepackage{xcolor}
\usepackage{booktabs}
\usepackage{graphicx}
\usepackage{comment}
\usepackage{tikz}
\usetikzlibrary{calc}

\newcommand{\x}{$\times$}

\newcommand{\squishlist}{
   \begin{list}{$\bullet$}
    { \setlength{\itemsep}{0pt}      \setlength{\parsep}{0pt}
      \setlength{\topsep}{3pt}       \setlength{\partopsep}{0pt}
      \setlength{\listparindent}{-2pt}
      \setlength{\itemindent}{-5pt}
      \setlength{\leftmargin}{1em} \setlength{\labelwidth}{0em}
      \setlength{\labelsep}{0.5em} } }

\newcommand{\squishend}{
    \end{list}  }

\newcommand*{\MinNumber}{20}%
\newcommand*{\MaxNumber}{100}%

\newcommand*\blackcircledempty[1]{\tikz[baseline=(char.base)]{
        \node[shape=circle, text={rgb,255:red,0;green,0;blue,0}, font=\small, draw={rgb,255:red,0;green,0;blue,0},inner sep=0.5pt] (char) {#1};}}  
        
\newcommand*\blackrectangleempty[1]{\tikz[baseline=(char.base)]{
        \node[shape=rectangle, text={rgb,255:red,0;green,0;blue,0}, font=\small, draw={rgb,255:red,0;green,0;blue,0},inner sep=0.6pt] (char) {#1};}}

\def\BibTeX{{\rm B\kern-.05em{\sc i\kern-.025em b}\kern-.08em
    T\kern-.1667em\lower.7ex\hbox{E}\kern-.125emX}}

\newcommand{\ApplyGradient}[1]{%
  \pgfmathsetmacro{\PercentColor}{(#1)/100.0*(\MaxNumber-\MinNumber) + \MinNumber}%
  \edef\x{\noexpand\cellcolor{black!\PercentColor}}\x\textcolor{white}{#1}%
}
\newcolumntype{R}{>{\collectcell\ApplyGradient}{r}<{\endcollectcell}}
\newcolumntype{C}{>{\collectcell\ApplyGradient}{c}<{\endcollectcell}}

\SetArgSty{textup}
\begin{document}
\title{Multi-node Acceleration for Large-scale GCNs}
\author{Gongjian Sun,
        ~Mingyu Yan,
        ~Duo Wang,
        ~Han Li,
        ~Wenming Li, \\
        ~Xiaochun Ye, 
        ~Dongrui Fan,~\IEEEmembership{Senior Member,~IEEE}
        ~and~Yuan~Xie,~\IEEEmembership{Fellow,~IEEE}%

\IEEEcompsocitemizethanks{
\IEEEcompsocthanksitem G. Sun, M. Yan, D. Wang, H. Li, W. Li, X. Ye, D. Fan are with the Institute of Computing Technology, Chinese Academy of Sciences, Beijing, China (Email: \{sungongjian19s, yanmingyu, wangduo18z, lihan-ams, liwenming, yexiaochun, fandr\}@ict.ac.cn). G. Sun, M. Yan, D. Wang, H. Li, W. Li, D. Fan are also with University of Chinese Academy of Sciences, Beijing, China.
Y. Xie is with University of California, Santa Barbara, California, USA (Email: yuanxie@ece.ucsb.edu). Corresponding author is Mingyu Yan (yanmingyu@ict.ac.cn).

}% <-this % stops an unwanted space
%\thanks{Corresponding author is Mingyu Yan (yanmingyu@ict.ac.cn).}
}

\markboth{To appear in TC}%
{Gongjian Sun \MakeLowercase{\textit{et al.}}: Multi-node Acceleration for Large-scale GCNs}

\IEEEtitleabstractindextext{%
\begin{abstract}
Limited by the memory capacity and compute power, singe-node graph convolutional neural network (GCN) accelerators cannot complete the execution of GCNs within a reasonable amount of time, due to the explosive size of graphs nowadays. 
Thus, large-scale GCNs call for a multi-node acceleration system (MultiAccSys) like TPU-Pod for large-scale neural networks. 

In this work, we aim to scale up single-node GCN accelerators to accelerate GCNs on large-scale graphs.
We first identify the communication pattern and challenges of multi-node acceleration for GCNs on large-scale graphs.
We observe that (1) coarse-grained communication patterns exist in the execution of GCNs in MultiAccSys, which introduces massive amount of redundant network transmissions and off-chip memory accesses; (2) overall, the acceleration of GCNs in MultiAccSys is bandwidth-bound and latency-tolerant.
Guided by these two observations, we then propose MultiGCN, the first MultiAccSys for large-scale GCNs that trades network latency for network bandwidth. 
Specifically, by leveraging the network latency tolerance, we \emph{first} propose a topology-aware multicast mechanism with a one \texttt{put} per \texttt{multicast} message-passing model to reduce transmissions and alleviate network bandwidth requirements. \emph{Second}, we introduce a scatter-based round execution mechanism which cooperates with the multicast mechanism and reduces redundant off-chip memory accesses. 
Compared to the baseline MultiAccSys, MultiGCN achieves 4$\sim$12$\times$ speedup using only 28\%$\sim$68\% energy, while reducing 32\% transmissions and 73\% off-chip memory accesses on average. 
It not only achieves 2.5$\sim$8$\times$ speedup over the state-of-the-art multi-GPU solution, but also scales to large-scale graphs as opposed to single-node GCN accelerators.
\end{abstract}

}

\maketitle

\IEEEdisplaynontitleabstractindextext

\IEEEpeerreviewmaketitle

\IEEEraisesectionheading{\section{Introduction}\label{sec:introduction}}

\IEEEPARstart{G}{raph} Convolutional Neural Networks (GCNs) have emerged as a premier paradigm to address the graph learning problem via generalizing the information encoding to graph topologies that can represent extremely complicated relationships~\cite{AliGraph,gnn_survey,gnn_review,graph_learning_survey,geometric_learning}. 
In reality, GCNs have been widely applied in many critical fields such as knowledge inference~\cite{gcn_knowledge_graph}, recommendation system \cite{gcn_recommender_systems}, visual reasoning \cite{gcn_visual_reasoning}, traffic prediction \cite{google_map_gnn}, Electronic design automation (EDA) \cite{GCN_EDA}, and GCN workloads can be found at many data centers \cite{AliGraph,google_map_gnn}.

GCNs typically exhibit a hybrid execution pattern introduced by two distinct execution phases, which hinder the acceleration of GCNs in GPUs~\cite{HyGCN,GCN_Characterization_CAL}.
The \texttt{Aggregation} phase traverses all vertices and aggregates the feature vectors of neighboring vertices into the current vertex, displaying an irregular execution pattern like graph processing (GP). The \texttt{Combination} phase further transforms the feature vector of each vertex into a new one using a multi-layer perceptron (MLP), exhibiting a regular execution pattern like neural network (NN). 
Such interleaving execution patterns hinder the acceleration of GCNs in GPUs which are inherently optimized for compute-intensive workloads with regular execution pattern~\cite{NVIDIA_Tesla}.

To tackle this hybrid execution pattern, previous efforts~\cite{HyGCN,GraphACT,AWB-GCN,EnGN,GCNAX} propose a series of single-node GCN accelerators.
Although these accelerators have achieved great improvement on both performance and energy efficiency compared with GPUs, they suffer from two following inefficiencies in the processing of large-scale GCNs. 
\textit{First}, with limited hardware resources from computation and buffering, a single-node accelerator cannot complete the execution of large-scale GCNs within a reasonable amount of time, not to mention that the sizes of real-world graphs keep growing rapidly~\cite{OGB,AliGraph}. 
\textit{Second}, massive time and energy have to be taken to move data between memory and hard disks~\cite{AliGraph}, because single-node accelerators do not have enough memory to accommodate the entire large-scale graph.
Thus, a multi-node acceleration system (MultiAccSys) is highly desirable for large-scale GCNs.

\begin{figure}[!t] 
    \centering
    \includegraphics[page=2,width=0.48\textwidth]{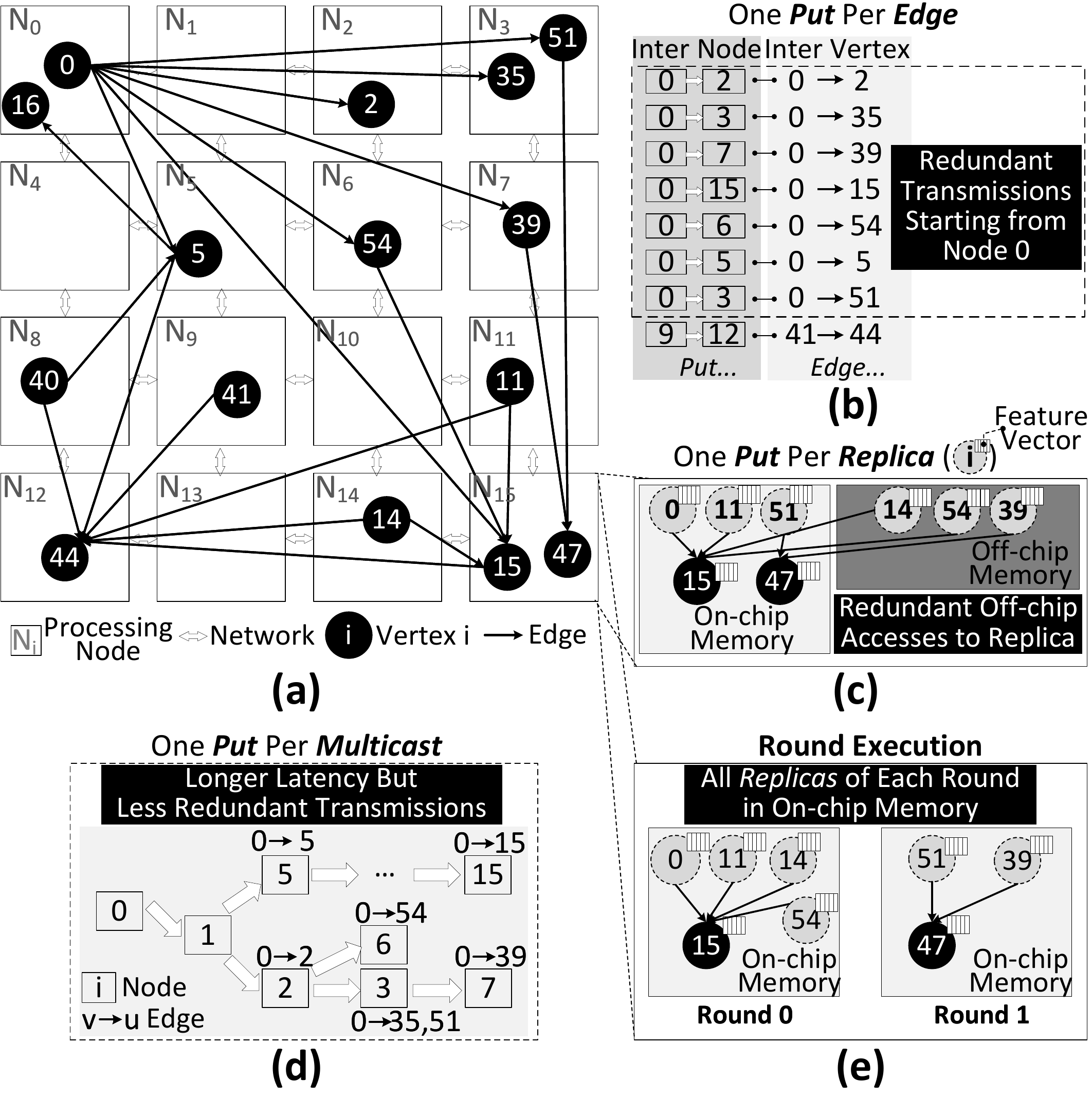} 
    \vspace{-8pt}
    \caption{
    Comparison between MultiGCN and previous efforts on MultiAccSys for graph processing: 
    (a) Examples of graph and MultiAccSys for GCNs; 
    (b) Disadvantage of MultiAccSys with one \texttt{put} per \texttt{edge} message-passing model; 
    (c) Disadvantage of MultiAccSys with one \texttt{put} per \texttt{replica} message-passing model;
    (d) and (e) Advantage of MultiGCN based on one \texttt{put} per \texttt{multicast} message-passing model and scatter-based round execution.}
    \label{fig:running_example}
\end{figure}

Previous efforts have proposed a series of MultiAccSyses for NNs and GPs, achieving great improvements on both performance and energy efficiency. However, they fail to tackle the unique execution pattern in the multi-node acceleration of GCNs.
MultiAccSyses for NNs, e.g., TPU-Pod~\cite{TPU-Pod}, tailor their hardware datapaths to the regular execution pattern such as coarse-grained regular communication pattern inter-node, to leverage the regularity for high performance of large-scale NN acceleration. 
Similarly, MultiAccSyses for GPs, e.g., Tesseract~\cite{Tesseract}, tailor their hardware datapaths to the irregular execution pattern of GPs such as fine-grained irregular communication pattern inter node, to alleviate the irregularity for high performance of large-scale GP acceleration.
Unfortunately, the multi-node acceleration for GCNs exhibits distinct execution patterns, i.e., the coarse-grained irregular communication inter node and hybrid execution pattern intra node.
Such execution patterns make MultiAccSyses ill-suited for GCNs, since they are only designed to either exploit regular execution pattern or alleviate irregular execution pattern.

In this paper, we aim to scale the single-node accelerator to accelerate GCNs on large-scale graphs like TPU-Pod~\cite{TPU-Pod}, since the performance and energy efficiency of single-node GCN accelerators are significantly higher than that of high-end GPUs.
To identify the unique execution pattern and the challenges of the multi-node acceleration for GCNs, we characterize the execution of large-scale GCNs in a straightforward design of MultiAccSys for GCNs as shown in Figure~\ref{fig:running_example}(a). To ensure an effective characterization, we borrow a well-designed single-node GCN accelerator~\cite{HyGCN} to design the processing node and a representative message-passing model to alleviate the irregular communication pattern inter node, i.e., one \texttt{put} per \texttt{edge} (OPPE)~\cite{Tesseract}. 
We observe irregular coarse-grained communication patterns in the execution of GCNs in MultiAccSys, which introduce massive amount of redundant network transmissions and off-chip memory accesses. 
This communication pattern is the direct result of the following facts: (1) each transmission between nodes contains a long feature vector of neighbor vertex and 
(2) it is unpredictable when and to where it needs to be sent due to the irregular connection pattern of neighbors. 
As a result, the OPPE message-passing model invokes many redundant coarse-grained transmissions because the long feature vector of each vertex must be repeatedly sent to all of its neighboring vertices. However, many of them may be sent to or pass through the same processing node. 
For example, in Figure~\ref{fig:running_example}(b), the feature vector of vertex V$_0$ in the processing node N$_0$ is sent to neighboring vertices V$_2$, V$_{35}$, V$_{51}$, V$_{39}$... in processing nodes N$_2$, N$_{3}$, N$_{3}$, N$_{7}$..., respectively. 
To reduce these redundant transmissions, a one \texttt{put} per \texttt{replica} (OPPR) message-passing model is proposed~\cite{GraphP}, which only puts one replica of the feature vector to each processing node and shares it with all neighboring vertices in that processing node. 
However, it is difficult to store thousands of replicas on-chip, which inevitably leads to massive amount of off-chip memory accesses.
For example, Figure~\ref{fig:running_example}(c) shows that vertices V$_{15}$ and V$_{47}$ in the processing node N$_{15}$ require accesses to the replicas of V$_{14}$, V$_{54}$, and V$_{39}$, which were stored off-chip after receiving because on-chip memory had been exhausted.

In light of the above challenges, we propose MultiGCN, the first MultiAccSys for GCNs, to accelerate the inference phase of large-scale GCNs by trading network latency for network bandwidth. 
\textit{\textit{First}, a topology-aware multicast mechanism with a one \texttt{put} per \texttt{multicast} message-passing model is proposed to alleviate network bandwidth requirements.} 
As shown in Figure~\ref{fig:running_example}(d), by leveraging the latency tolerance we identified as well as the known network and graph topologies, V$_0$'s feature vector is multicast in the transmission to reduce redundant transmissions.
\textit{\textit{Second}, a scatter-based round execution mechanism is proposed to cooperate with the multicast mechanism, which inherently matches the behaviour of multicast.} Specifically, each processing node scatters the replicas of vertices' feature vectors to remote processing nodes which perform aggregation for their neighboring vertices. 
Besides, to reduce redundant off-chip memory accesses, the graph is partitioned into a certain number of sub-graphs, one for each execution round, as shown in Figure~\ref{fig:running_example}(e). Thus, all replicas of round 0 (i.e., V$_0$, V$_{11}$, V$_{14}$, and V$_{54}$) and round 1 (i.e., V$_{51}$ and V$_{39}$) from remote processing nodes can be stored on-chip until the corresponding round completes. 
Intra- and inter-round overlap are utilized for higher performance.

The key contributions of this paper are as follows: \par

\squishlist

    \item We identify the communication pattern and challenges of multi-node acceleration for large-scale GCNs and observe that:
    (1) There are irregular coarse-grained communication patterns that invoke massive redundant transmissions and off-chip memory accesses; (2) Execution of GCNs in MultiAccSys is mainly bandwidth-bound and latency-tolerant.

    \item Accordingly, we propose the first multi-node acceleration system for GCNs which scales single-node GCN accelerators to accelerate GCNs on large-scale graphs by trading network latency for network bandwidth. 
    
    \item We propose a topology-aware multicast mechanism to reduce redundant transmissions and  alleviate network bandwidth requirements, along with a scatter-based round execution mechanism that cooperates with the multicast mechanism to reduce off-chip memory accesses.

    \item We implement MultiGCN in both RTL and cycle-accurate simulator to demonstrate its advantages.
    Compared with OPPE-based MultiAccSys, MultiGCN achieves 4$\sim$12$\times$ speedup, and reduces 32\% network transmissions as well as 73\% off-chip memory accesses on average. 
    Besides, MultiGCN achieves 2.5$\sim$8$\times$ speedup over the multi-GPU solution and can scale to large-scale graph as opposed to single-node GCN accelerators.

    \item  We explore the design of MultiGCN in detail and provide insights into the relationships between hardware parameters, graph characteristics, and architecture techniques.
    
\squishend

\begin{table}[!htbp]
    \vspace{-8pt}
	\caption{Notations used in this paper.} \label{tab:notation}
    \vspace{-8pt}
	\centering
	\renewcommand\arraystretch{1.0}
    \resizebox{0.43\textwidth}{!}{
    \begin{tabular}{c|c}
    \hline
        \textbf{Notation} & \textbf{Explanation} \\ \hline
        $G=(V,E)$ & directed graph $G$ \\
        $V$($|V|$) & (size of) vertex set of graph $G$ \\
        $E$($|E|$) & (size of) edge set of graph $G$ \\
        $(i,j)$ or $e_{i,j}$ & edge from vertex $i$ to vertex $j$ \\
        $d_v$ & in-degree of vertex $v$ \\
        $N_v$ & incoming neighbor set of vertex $v$ \\
        $h_v^k$($|h_v^k|$) & (length of) feature vector of vertex $v$ at $k$-th layer \\
        $a_v^k$ & aggregated result of vertex $v$ at $k$-th layer \\
        \hline
    \end{tabular}
    	}
\end{table}

\vspace{-15pt}

\begin{figure}[!htbp] 
    \centering
    \includegraphics[page=10, width=0.8\linewidth]{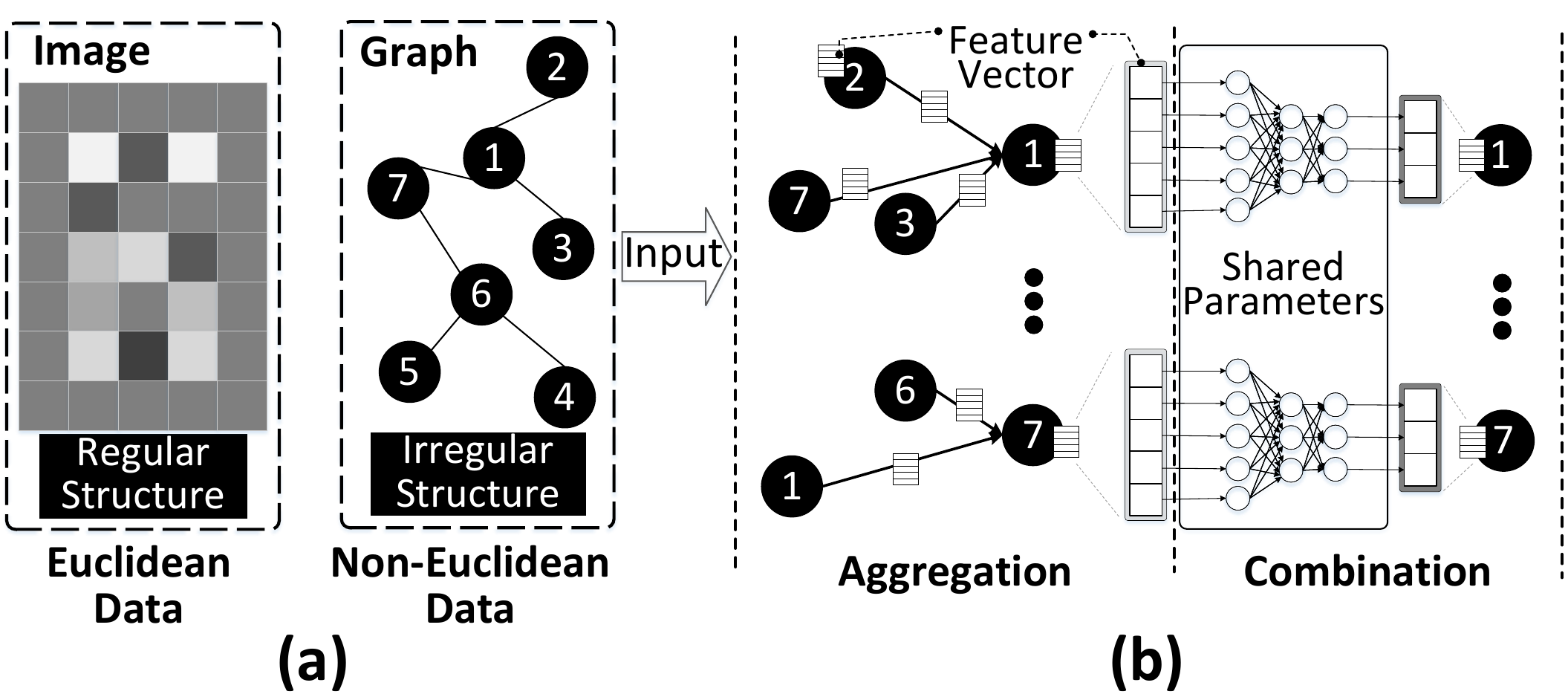}
    \vspace{-8pt}
    \caption{Illustration examples of (a) graph and (b) GCNs execution.}
    \label{fig:graph_GCN}
\end{figure}

\begin{figure*}[!t] 
    \centering
    \includegraphics[page=1, width=1\linewidth]{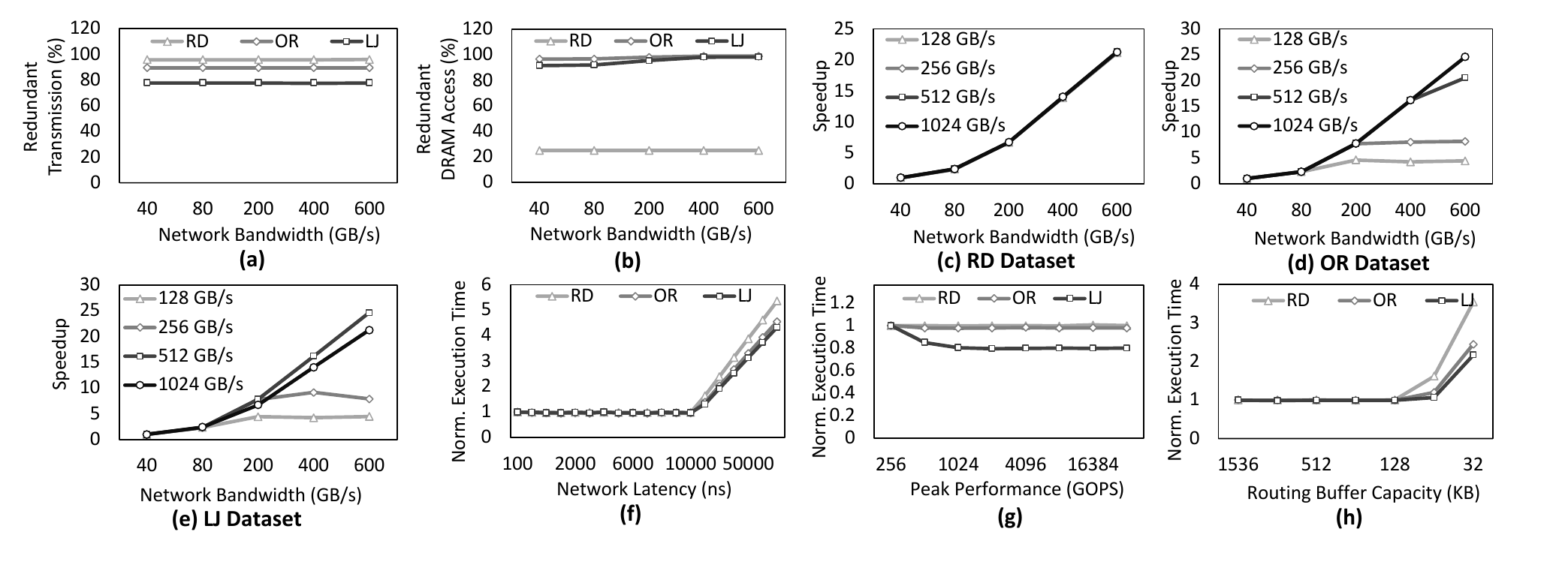}
    \vspace{-15pt}
    \caption{Results of characterization on OPPE-Based MultiAccSys with 16 processing nodes:
     (a) Ratio of redundant transmissions to total transmissions across different network bandwidths; (b) Ratio of redundant DRAM accesses to total DRAM accesses across different network bandwidths; Speedup of GCN across different network bandwidths (X-axis) and DRAM bandwidths (4 Lines) on (c) RD, (d) OR, and (e) LJ datasets; Normalized execution time across (f) different network latencies, (g) different peak performances, and (h) different routing buffer capacities.}
    \label{fig:MultiAccSys_characterizations}
\end{figure*}

\section{Background} \label{sec:background} 

\textbf{GCNs.} 
As shown in Figure~\ref{fig:graph_GCN}, typical GCNs take non-euclidean
data as input, i.e., graphs which have irregular structure and differ from the grid-structured images as shown in Figure~\ref{fig:graph_GCN}(a). GCNs typically consist of several graph convolutional layers, each with two main phases: \texttt{Aggregation} and \texttt{Combination}, which can be formulated as:
\[a_v^k=\underset{u\in \{v\}\cup N_v}{Aggregate}\left(h_u^{k-1}\right), h_v^k=Combine\left(a_v^k\right),\]
where $v\in V$. For clarity, by ``node'' we refer to the processsing nodes in MultiAccSys, and by ``vertex'' we refer to the elements of the graph vertex set. 

In the $k$-th layer, for each vertex $v$, the feature vectors $h_u^{k-1}$ of neighboring vertices in $N_v$ are aggregated into $a_v^k$
% following the graph topology by \texttt{aggregate} function in \texttt{Aggregation} phase
, as shown in Figure~\ref{fig:graph_GCN}(b).
Since \textit{Aggregation} phase heavily relies on the graph structure that can be arbitrary and sparse, it suffers from irregular data communication patterns.
The \texttt{combine} function further transforms $a_v^k$ to generate $h_v^k$ using an MLP in the \texttt{Combination} phase. All vertices perform \texttt{combine} function using a shared MLP,
which results in intensive computation and high data locality in the \textit{Combination} phase.
The feature vector $h_v^{k}$ can be very long, and often up to thousands in the initial feature vectors $h_v^0$ of all vertices.

\textbf{Network and Message-passing Model.} 
Multi-node acceleration becomes an inevitable trend due to the ever-increasing demand on computation and storage capability in emerging domains such as deep learning~\cite{NVLink}. 
High-speed network interfaces and well-designed network topologies are the basics of effective multi-node acceleration. 
Network interfaces include NVLink for GPU interconnection, PCIe for CPU and GPU interconnection, and so on. 
Network latency is a metric used to evaluate how much time is consumed by the network interface to send a minimal packet from the source to the destination.
Network topology includes star, bus, mesh, torus, and so on. 
For example, a NVIDIA HGX houses 16 H100 GPUs networked together using NVLink and NVSwitch in star topology \footnote{https://www.nvidia.cn/data-center/hgx/}, with a peak communication bandwidth of up to 900GB/s. 
The message-passing model determines the transfer mode of MultiAccSys, such as the OPPE and OPPR models used in the MultiAccSys of GPs. For each vertex, the OPPE model sends one replica of the feature vector to each neighboring vertex, while the OPPR model only sends one replica of the feature vector to each processing node and shares it with all neighboring vertices in that processing node.

\section{Motivation} \label{sec:motivation}

\textbf{Inefficiencies of Single-node GCN Accelerators.}
Previous efforts propose several single-node accelerators for GCN acceleration, achieving significant improvements in both performance and efficiency compared with GPUs. 
However, the ever-growing scale of graphs hinders efficient execution of GCNs on single-node accelerators. 
For example, to tackle the hybrid execution pattern, HyGCN~\cite{HyGCN} proposes a hybrid architecture for GCNs, which includes two engines respectively tailored to \textit{Aggregation} and \textit{Combination} phases. HyGCN achieves an average 6.5$\times$ speedup with 10$\times$ energy reduction over high-end GPUs.
However, with limited off-chip memory bandwidth, on-chip buffer, and compute resources, a single-node accelerator cannot process large-scale GCNs within a reasonable amount of time, while the scale of real-world graphs contines to grow rapidly~\cite{OGB,AliGraph}.
In addition, large-scale graphs demand massive memory, which is hard to satisfy within a single-node accelerator~\cite{AliGraph}, resulting in costly data transfers between the memory and the hard disk.
Thus, a MultiAccSys for GCNs is highly desirable.

\textbf{Inefficiencies of NN and GP MultiAccSyses.} 
Previous efforts propose a series of MultiAccSyses for large-scale NNs and GPs.
However, they fail to tackle the unique execution pattern of the multi-node acceleration for GCNs.
For example, the designers of TPU-Pod~\cite{TPU-Pod} elaborately customize an MultiAccSys for NNs using an NVLink-like inter-node network interface to connect TPU chips.
Although TPU-Pod delivers near-linear speedup for large-scale NNs, 
they are
%elaborately tailored to the regular coarse-grained communication pattern and regular execution pattern, and hence 
ill-suited for GCNs due to the irregular coarse-grained communication pattern and hybrid execution patterns. 
Another example is Tesseract~\cite{Tesseract}, whose designers elaborately customize an MultiAccSys for GPs using a well-designed message passing model (i.e., OPPE) to alleviate the fine-grained irregular communication pattern. 
Although Tesseract achieves significant speedup for large-scale GPs, it invokes massive amounts of redundant coarse-grained transmissions caused by the irregular coarse-grained communication pattern. 
Although the redundant transmissions can be reduced by introducing the OPPR message-passing model~\cite{GraphP}, the replicas of feature vectors can be prohibitively large to be stored entirely on-chip, which inevitably leads to many off-chip memory accesses.

\textbf{Characterization on A Straightforward Design.} 
To identify the communication pattern and challenge of the multi-node acceleration for GCNs, 
a detailed characterization is conducted on an OPPE-based MultiAccSys and results are shown in Figure~\ref{fig:MultiAccSys_characterizations}.
Here, we briefly introduce the experimental setup. Please see Section~\ref{sec:Methodology} for our detailed evaluation methodology.
We use a OPPE-based MultiAccSys consisting of 16 processing nodes, connecting in a manner of $4\times 4$-sized 2D torus bidirectional network. Each processing node is a variant of the single-node GCN accelerator of previous work~\cite{HyGCN}, having exactly the same hardware resources and one disjoint part of the graph data.
The message-passing model used in this MultiAccSys is inspired by the OPPE model, which aims to tackle the irregular communication pattern caused by the irregular graph topology. 
An in-house cycle-accurate simulator is designed and implemented to measure execution time. We additionally implement the core modules in Verilog and synthesized using Synopsys compilers to check clock frequency requirement and estimate performance metrics. 

We observe that the irregular coarse-grained communication patterns exist in the execution of GCNs in MultiAccSys, which introduce massive amounts of redundant network transmissions and off-chip memory accesses. 
The irregular coarse-grained communication pattern is caused by two reasons:
(1) each transmission between node contains a long feature vector of neighbor vertex, with up to hundreds of elements, determined by the size of the input dataset or the number of the MLP's output neurons, and 
(2) it is unpredictable when and to where a long feature vector needs to be sent due to the irregular connection pattern of neighbors in graph. 
As depicted in Figure~\ref{fig:MultiAccSys_characterizations}(a) and (b), we observe a vast range of redundant transmissions and DRAM accesses, ranging from 78\% to 96\% and 25\% to 99.9\%, respectively. 
This is because the long feature vectors of each vertex must be repeatedly sent to all of its neighboring vertices, many of which may be sent to or through the same processing nodes.  
In addition, received feature vectors need to be saved to DRAM upon receipt and loaded from DRAM while in use due to the large number of long feature vectors and limited on-chip cache capacity.
These redundancies waste network bandwidth and DRAM bandwidth, which significantly hinders the performance and efficiency on MultiAccSys for GCNs.

We also observe that the acceleration of GCNs in MultiAccSys is bandwidth-bound and latency-tolerant. 
Figure~\ref{fig:MultiAccSys_characterizations}(c), (d), and (e) show that the speedup across different datasets grows almost linearly as network bandwidth increases when DRAM bandwidth is sufficient (i.e., greater than 256 GB/s).
This is because neighboring feature vectors with hundreds of elements for each vertex need to be sent and aggregated in a target processing node, following the irregular neighbor connections in \textit{Aggregation} phase, which consumes much network bandwidth for better performance. 
However, Figure~\ref{fig:MultiAccSys_characterizations}(f) shows that the normalized execution time is nearly constant under different network latencies until around 20,000 \textit{ns}. 
This value mainly relies on the processing time which is positively correlated with the length of the feature vector and negatively correlated with the DRAM bandwidth of the processing node.
This is because the received feature vectors is frequently stored to or load from DRAM as aforementioned.
Figure~\ref{fig:MultiAccSys_characterizations}(g) shows that the normalized execution time is nearly constant under different peak performances with more than 1024 giga operations per second (GOPS).
This is because the low utilization of network bandwidth and DRAM bandwidth become the performance bottleneck.
Figure~\ref{fig:MultiAccSys_characterizations}(h) shows that the normalized execution time is also nearly constant under different routing buffer capacities above 64 KB.
Since the routing buffer is used to buffer the routing packets before they are sent, 
its capacity relies on the utilized network bandwidth and network latency.

\section{MultiGCN Architecture} 

Guided by the above observations, we propose MultiGCN, an efficient MultiAccSys for large-scale GCNs that trades network latency for network bandwidth.

\subsection{Architecture Overview}

\begin{figure*}[!t]
    \centering
    \includegraphics[page=3,width=0.8\textwidth]{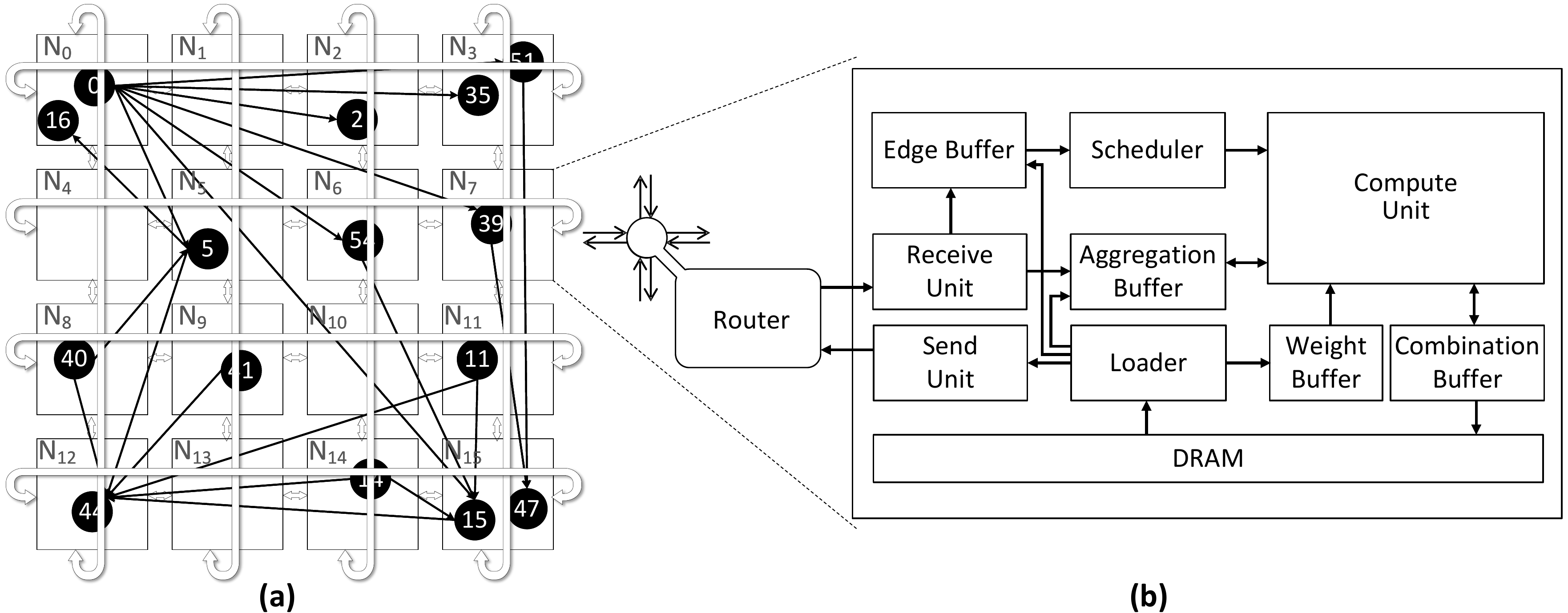} 
    \vspace{-8pt}
    \caption{Architecture of MultiGCN: (a) 2D torus network with 16 processing nodes; (b) Design of processing node.}
    \label{fig:arch-overview}
    \vspace{-8pt}
\end{figure*}

Figure~\ref{fig:arch-overview} provides an illustration of the proposed architecture. Noting that MultiGCN does not rely on a particular network topology, we choose a 2D torus topology consisting of 16 processing nodes as our baseline, which is shown in Figure~\ref{fig:arch-overview}(a). 
For network links we use the NVLINK protocol, which is one of the most widely used high-speed interconnection protocols and usually used between NVIDIA GPUs. 
A processing node, shown in Figure~\ref{fig:arch-overview}(b), is composed of a compute unit, a router, a receive unit, a send unit, an edge buffer, a scheduler, a loader, an aggregation buffer, a weight buffer, a combination buffer, and DRAM. 

The compute unit consists of eight reusable 1$\times$128 systolic arrays. Each processing element (PE) has four registers with two for input, one for output, and one for internal use respectively, and an ALU capable of multiplication and reduction (like MIN, MAX, ADD). The eight systolic arrays work separately, either in combination mode, like a traditional systolic array, or in aggregation mode. 
In aggregation mode, all PEs follow an identical three-stage pipeline: read two operands from the input registers, perform reduction, and write the result to the output register. Moreover, a real-time scheduling of compute resources between aggregation and combination is implemented in MultiGCN, since all eight reusable systolic arrays can process workloads of both types.
Note that although a unified compute unit is used in this work, designs in other single-node accelerators can also be integrated for better efficiency or compatibility.

The router, receive unit, and send unit are used to transfer vertices' feature vectors and neighboring lists. 
The edge buffer and scheduler are used to efficiently organize computation. Each entry in the edge buffer contains the address of a vertex's feature vector in aggregation buffer and its neighbor list. 
The feature vector is read via address and aggregated into the intermediate result of vertices in the neighbor list. The aggregation process is recorded aside the intermediate result in the aggregation buffer.
The weight buffer and combination buffer save the weight matrix and intermediate result for the combination process.
The loader loads the meta-data of execution, ID and degree of vertices, feature vectors and edge lists, which the send unit and scheduler ingest to complete execution.

\textbf{Topology-aware Multicast (Section~\ref{sec:multicast}).} 
To reduce the requirement of network bandwidth, a topology-aware multicast mechanism with a one \texttt{put} per \texttt{multicast} message-passing model is proposed. 
Multicast is based on the simple idea that for a vertex $v$, many processing nodes where $v$'s neighbors lie in can be satisfied by receiving the same packet containing a replica of $v$'s feature vector.
To efficiently tailor multicast to the communication pattern of GCNs, we design our multicast to be network topology-aware and graph topology-aware. The network topology awareness helps route and split packets guided by routers’ status and an explicit list of destination node IDs in the packet. Thus, the packet can be quickly and exactly multicast to all destination nodes.
The graph topology awareness is enabled by the offset list and neighbor lists in the packet which are used to share exactly one replica to all neighbors in the same processing node.
Although these capabilities introduce extra latency during transmission because of the above additional info in the packets, they help eliminate redundant transmissions and significantly reduce the need for network bandwidth.

\textbf{Scatter-based Round Execution (Section~\ref{sec:round_execution_mechanism}).} 
Although the topology-aware multicast mechanism helps reduce the requirements of network bandwidth, its overhead is high for three reasons. 
\textit{First}, a request-response loop, required by each replica’s transmission, significantly increases the design complexity of MultiAccSys for the multicast mechanism. 
\textit{Second}, the large offset list and neighbor lists are coupled into a single packet, resulting in an unacceptable routing latency in multicast. 
\textit{Third}, on-chip memory is unable to buffer all thousands of replicas of long feature vectors, which leads to frequent replacements of replicas between on-chip memory and off-chip memory.
To this end, a scatter-based round execution mechanism is proposed, which inherently matches the behaviour of multicast. 
Specifically, each processing node scatters the replicas of vertices’ feature vectors to remote processing nodes who perform the \texttt{aggregate} function for these vertices’ neighboring vertices, so that the request-response loop is eliminated. 
Besides, the graph is partitioned into multiple sub-graphs, one for each execution round. 
Thus, the large neighbor lists are partitioned and transmitted over several rounds, avoiding the unacceptable routing latency. Moreover, only a small number of replicas in each round are stored on-chip until the corresponding round completes, reducing redundant off-chip memory accesses.

\subsection{Topology-aware Multicast Mechanism} \label{sec:multicast}
To enable efficient multicast, 
we propose a one \texttt{put} per \texttt{multicast} message-passing model based on the DyXY routing algorithm~\cite{DyXY_routing_method} to implement the multicast mechanism for a given network topology and graph topology. 

\textbf{One Put per Multicast.} 
Our one \texttt{put} per \texttt{multicast} model is inspired by multicast on Ethernet. In Ethernet, multicast is a group communication where data transmission is addressed to a group of destination computers simultaneously. 
In MultiGCN, \texttt{put} refers to putting the replica of a vertex's feature vector to its neighboring vertices, while \texttt{multicast} means the packet generated by \texttt{put} is scattered to all its destinations in Ethernet multicast-like fashion. 

\begin{figure*}[!htbp]
    %\vspace{-5pt}
    \centering
    \includegraphics[page=5,width=0.92\textwidth]{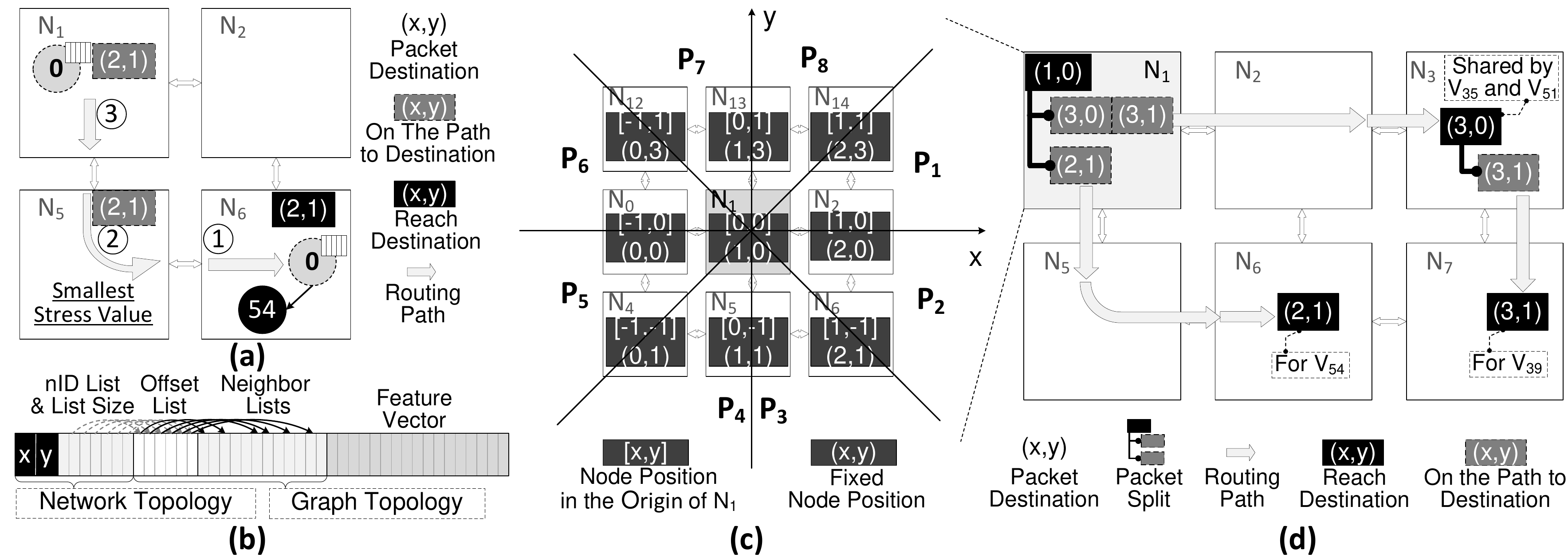} 
    \vspace{-8pt}
    \caption{Illustration of topology-aware multicast based on DyXY routing algorithm: (a) Example of DyXY routing algorithm; (b) Packet format for topology-aware multicast; (c) Packet split of topology-aware multicast in the step 1 of DyXY; (d) One possible multicast of V$_0$'s feature vector from N$_1$ (Origin) to V$_{35}$ and V$_{51}$ in N$_3$, V$_{39}$ in N$_7$, and V$_{54}$ in N$_6$.}
    \label{fig:optimization_routing}
    %\vspace{-15pt}
\end{figure*}

The basic routing algorithm we use is DyXY \cite{DyXY_routing_method}, which provides adaptive routing based on congestion status nearby. The algorithm is adaptive in that it makes routing decisions by monitoring the congestion status nearby.
It is also deadlock-free and livelock-free since the path of a packet in the network to one of the shortest paths between the source and the destination is limited.
If multiple shortest paths are available, the routers will help the packet to choose one of them based on the congestion condition of the network. A stress value, i.e., the occupation ratio of the routing buffer, is used to represent the congestion condition of a router in this work.
Each router stores instant stress values for all neighbors, which are updated periodically. The detailed routing algorithm is shown in Algorithm \ref{alg:DyXY_routing} and a routing example for the replica of V$_0$'s feature vector is depicted in Figure~\ref{fig:optimization_routing}(a). See Section~\ref{sec:round_execution_mechanism} for details of graph mapping.

\begin{algorithm}[!t]
 \small
 \label{alg:DyXY_routing}
  \caption{\textbf{DyXY Routing Algorithm}}
  \ForEach{packet~\texttt{p}~to~(x,y)~in~routing~buffer}{
  \blackcircledempty{1} \If{(x == Tx) \&\& (y == Ty)}
  {
    receive \texttt{p} in current node (Tx, Ty);   \\
  }
  \blackcircledempty{2} \ElseIf{(x == Tx) $\|$ (y == Ty)}{
    send \texttt{p} to neighbor on Y-axis or X-axis; \\
  }
  \blackcircledempty{3} \Else{
    send \texttt{p} to neighbor with smallest stress value; \\
  }
}
\end{algorithm}

To couple the multicast mechanism with the DyXY routing algorithm, step \blackcircledempty{1} in Algorithm \ref{alg:DyXY_routing} is modified to split packets as shown in Algorithm \ref{alg:optimization_routing}. 
Figure~\ref{fig:optimization_routing}(b) shows the packet format used in Algorithm \ref{alg:optimization_routing}, which consists of four parts including the position of the next destination node (x, y), network topology (neighbor-ID (nID) list and its size), graph topology (offset list, neighbor lists), and the replica of feature vector. 
In the line \blackrectangleempty{2} of Algorithm \ref{alg:optimization_routing}, all nIDs in the nID list are transformed into the new coordinate [x,y] by taking the current node (Tx, Ty) as the Origin of coordinates and donated as a set D = \{[x,y]\}, when a packet arrives at the destination node (Tx, Ty). 
Figure~\ref{fig:optimization_routing}(c) gives an example for the transformation of nID, new node position [x,y], and fixed node position (x,y). In this example, N$_1$ (1, 0) is taken as the Origin [0,0], and the [x,y] of its neighboring nodes are shown. 
In line \blackrectangleempty{3} to \blackrectangleempty{12}, based on the new coordinates, the packet is partitioned into nine parts.
Each part has a part of the nID list and neighbor lists, a new offset list, and a complete replica.
In line \blackrectangleempty{14} to \blackrectangleempty{41}, these nine parts are received by the current node or sent to the next destination node.
Figure~\ref{fig:optimization_routing}(d) gives a multicast example based on Algorithm~\ref{alg:optimization_routing} where the replica of V$_0$'s feature vector is multicast from N$_1$ to V$_{35}$ and V$_{51}$ in N$_3$, V$_{39}$ in N$_7$, and V$_{54}$ in N$_6$. Specifically, the packet arrives at N$_1$ (1,0) and is then split into two parts. 
One is P$_3$ and consists of one destination including [1, -1] (i.e., N$_6$ (2, 1)). 
The other one is P$_2$ and consists of two destination nodes including [2, 0] (i.e., N$_3$ (3, 0)) and [2, -1] (i.e., N$_7$ (3, 1)). 
Then, the former is sent to N$_6$ via N$_5$. 
The latter is sent to N$_3$ via N$_2$ and is further multicast until the nID list in packet is empty. The packet received by N$_3$ is further shared by the aggregation of V$_{35}$ and V$_{51}$ according to the neighbor list in the packet. 
As a result, redundant transmissions are eliminated.

In this process, we have followed the spirit of trading latency for network bandwidth: although the additional info in the packets induce extra transfer latencies, the topology-aware multicast mechanism alleviates the requirement of network bandwidth.
However, this also introduces three inefficiencies: high design complexity, low utilization of compute resource, and redundant off-chip memory accesses.
\textit{First}, a request-response loop, required by each transmission of the replica, will significantly increase the design complexity. 
\textit{Second}, the large neighbor lists in the packet cause intensive transmission and unacceptable routing latency, and most of the compute resources become underutilized. 
\textit{Third}, limited by the capacity of on-chip memory in each processing node, the large volume of received replicas need to be frequently moved between on-chip memory and off-chip memory. 
This is because thousands of replicas are received and shared by many vertices' aggregation in each processing node, but it takes a long time to process the long feature vectors, which means most of these replicas need to be first stored in off-chip memory and then reloaded when needed.

\begin{algorithm}[!t]
 \small
 \label{alg:optimization_routing}
  \caption{\textbf{Packet Split of Multicast}}
  \blackcircledempty{1} \If{(x == Tx) \&\& (y == Ty)}
  {
    transform nID list into D = \{[x,y]\} by taking node (Tx, Ty) as the Origin of coordinates; \\
    split packet \texttt{p} into the following nine parts: \\ 
    P$_0$ = \{[0, 0]\} $\cap$ D ; \\
    P$_1$ = \{[x, y] $|$ y $>$ 0, y $\leq$ ~x\} $\cap$ D ; \\
    P$_2$ = \{[x, y] $|$ y $\leq$ 0, y $>$ -x\} $\cap$ D ; \\
    P$_3$ = \{[x, y] $|$ x $>$ 0, y $\leq$ -x\} $\cap$ D ; \\
    P$_4$ = \{[x, y] $|$ x $\leq$ 0, y $<$ ~x\} $\cap$ D ; \\
    P$_5$ = \{[x, y] $|$ y $<$ 0, y $\geq$ ~x\} $\cap$ D ; \\
    P$_6$ = \{[x, y] $|$ y $\geq$ 0, y $<$ -x\} $\cap$ D ; \\
    P$_7$ = \{[x, y] $|$ y $\geq$ -x, x $<$ 0\} $\cap$ D ; \\
    P$_8$ = \{[x, y] $|$ x $\geq$ 0, y $>$ ~x\} $\cap$ D ; \\
    
    receive and share P$_0$ in current node (Tx, Ty);   \\
    
    \If{(P$_1$ $\neq$ $\varnothing$) \&\& (P$_2$ $\neq$ $\varnothing$)}
    {
        send P$_1$ $\cup$ P$_2$ to [MIN(P$_1$.x $\cup$ P$_2$.x), 0]; \\
    }
    \Else 
    {
        send P$_1$ to [MIN(P$_1$.x), MIN(P$_1$.y)]; \\
        send P$_2$ to [MIN(P$_2$.x), MAX(P$_2$.y)]; \\
    }
    
    \If{(P$_3$ $\neq$ $\varnothing$) \&\& (P$_4$ $\neq$ $\varnothing$)}
    {
        send P$_3$ $\cup$ P$_4$ to [0, MAX(P$_3$.y $\cup$ P$_4$.y)]; \\
    }
    \Else 
    {
        send P$_3$ to [MIN(P$_3$.x), MAX(P$_3$.y)]; \\
        send P$_4$ to [MAX(P$_4$.x), MAX(P$_4$.y)]; \\

    }
    
    \If{(P$_5$ $\neq$ $\varnothing$) \&\& (P$_6$ $\neq$ $\varnothing$)}
    {
        send P$_5$ $\cup$ P$_6$ to [MAX(P$_5$.x $\cup$ P$_6$.x), 0]; \\
    }
    \Else 
    {
        send P$_5$ to [MAX(P$_5$.x), MAX(P$_5$.y)]; \\
        send P$_6$ to [MAX(P$_6$.x), MIN(P$_6$.y)]; \\

    }
    
    \If{(P$_7$ $\neq$ $\varnothing$) \&\& (P$_8$ $\neq$ $\varnothing$)}
    {
        send P$_7$ $\cup$ P$_8$ to [0, MIN(P$_7$.y $\cup$ P$_8$.y)]; \\
    }
    \Else 
    {
        send P$_7$ to [MAX(P$_7$.x), MIN(P$_7$.y)]; \\
        send P$_8$ to [MIN(P$_8$.x), MIN(P$_8$.y)]; \\
    }
  }
\end{algorithm}

\subsection{Scatter-based Round Execution Mechanism} \label{sec:round_execution_mechanism}
To address the above inefficiencies, we propose a scatter-based round execution mechanism that well suits the topology-aware multicast mechanism. The key idea of the scatter-based round execution mechanism is simple but effective: we first partition the graph into a set of sub-graphs and then process one sub-graph per round. 
In each round, all replicas are kept on-chip until no longer needed.
To improve resource utilization, we also implement intra- and inter-round overlaps.

\begin{figure}[!t] 
    \vspace{-5pt}
    \centering
    \includegraphics[page=6,width=0.45\textwidth]{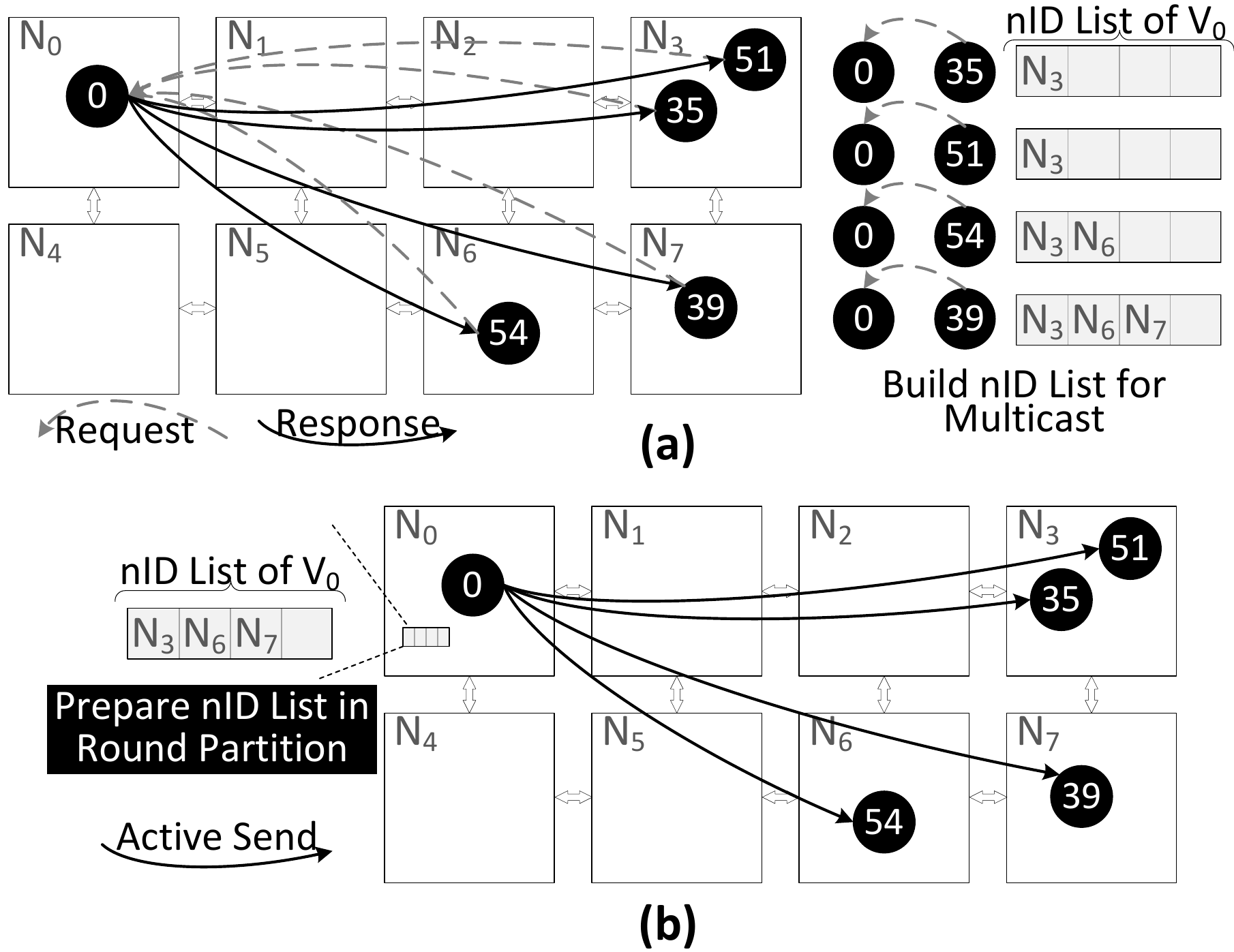} 
    \vspace{-8pt}
    \caption{The execution of \texttt{aggregate} function: (a) Gather-based method; (b) Scatter-based method.}
    \label{fig:round_parition_implement_method}
\end{figure}

\textbf{Scatter or Gather.} 
There are two ways to execute the \texttt{aggregate} function: gather-based and scatter-based methods. As shown in Figure~\ref{fig:round_parition_implement_method}(a), in the gather-based method, each processing node (e.g., N$_{3}$, N$_{6}$, and N$_{7}$) first requests feature vectors of neighboring vertices (e.g., V$_0$) for each vertex (e.g., V$_{35}$, V$_{51}$, V$_{54}$, and V$_{39}$) from the remote processing node (e.g., N$_{0}$) and then waits for the responded feature vectors. Afterwards, the replicas of feature vectors are aggregated in the processing node of the requester (e.g., N$_{3}$, N$_{6}$, and N$_{7}$). As a result, a request-response loop for each transmission of the replica is introduced. 
Besides, to support multicast, the remote processing node needs to collect requests for each vertex's feature vector to build an nID list and then performs multicast based on this nID list.

As shown in Figure~\ref{fig:round_parition_implement_method}(b), in the scatter-based method, each processing node has an nID list derived from round partition (described in the next paragraph). Each processing node (e.g., N$_{0}$) actively sends the feature vector of each vertex (e.g., V$_{0}$) to the remote processing nodes (e.g., N$_{3}$, N$_{6}$, and N$_{7}$) where the out-going neighboring vertices (e.g., V$_{35}$, V$_{51}$, V$_{54}$, and V$_{39}$) reside. 
Then, the feature vectors are aggregated in the remote processing nodes. As a result, message passing only happens in a single direction. From the above analysis, it can be seen that the scatter-based method inherently matches the behavior of topology-aware multicast mechanism, helping eliminate the request-response loop, and thus achieves lower design complexity than the gather-based method. Hence, we employ the scatter-based method for MultiGCN.

\begin{figure}[!t] 
    %\vspace{-5pt}
    \centering
    \includegraphics[page=7,width=0.48\textwidth]{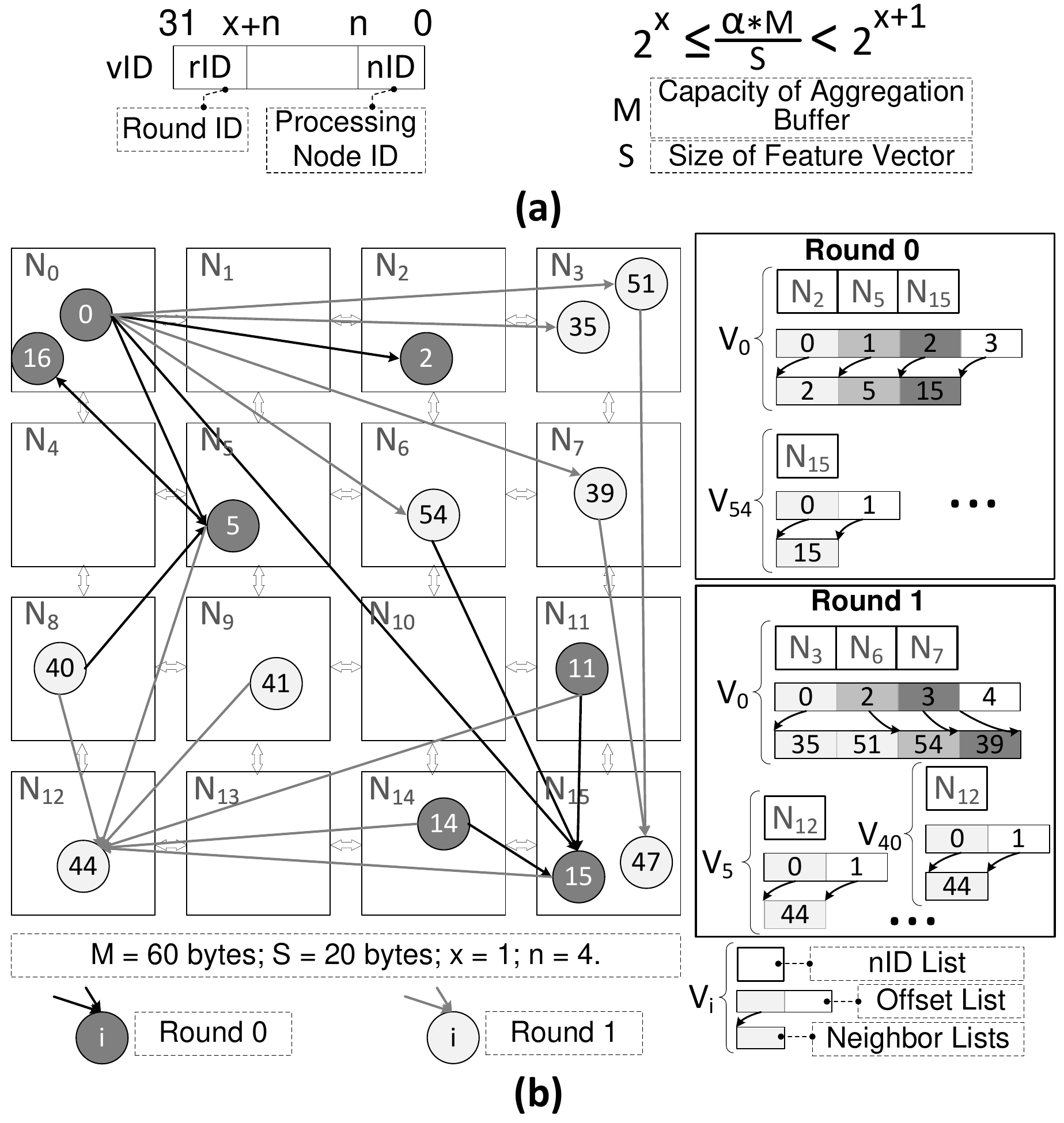} 
    \vspace{-8pt}
    \caption{Illustration of round partition: (a) Generation of round ID; (b) Example of round partition.}
    \label{fig:round_parition}
\end{figure}

\textbf{Round Partition and Execution.} 
To avoid unacceptable routing latency and redundant off-chip memory accesses, a round execution method coupled with a simple round partition is proposed. 
Figure~\ref{fig:round_parition} illustrates the round partition which is based on a simple graph mapping method to reduce the mapping overhead and simplify packet routing.
As illustrated in Figure~\ref{fig:round_parition}(a), for a vertex, 
the bits in range $[0, n)$ in the vertex ID (vID) will be the ID of the processing node to which the vertex is partitioned and mapped. 
The bits in range $[n, x+n)$ in the vID are used to partition and map $2^{x}$ vertices with interleaving vIDs into the same processing node together. 
The bits in range $[x+n, 32)$ in the vID will be the index of round (rID) for simplicity. 
The value of $n$ is determined by the total number of processing nodes ($\rm \#total\_node$) in MultiAccSys, which is equal to $\lfloor \log_2(\rm \#total\_node) \rfloor$.
The value of $x$ can be reconfigured for different datasets to better utilize on-chip memory, which is determined by $2^x\leq \frac{\alpha M}{S} < 2^{x+1}$, where M is the capacity of the aggregation buffer and S is the size of aggregated feature vector. 
The value of ${\alpha}$ must be less than 1 to reserve space for network communication and round overlap.
In our implementation, we set ${\alpha}$ to 0.75.
For each vertex, all its in-coming edges are partitioned into the same round, and used to build the nID list, offset list, and neighbor lists of its all in-coming neighboring vertices for multicast. The compressed sparse row format is used to reorganize the neighbor lists to reduce needs for both memory and network bandwidth. These information of each vertex is mapped into the same processing node with this vertex's feature vector. Note that after the round partition, if a vertex still has too many outgoing neighbors in a round, this packet is further divided into several packets before being sent to network.
Figure~\ref{fig:round_parition}(b) provides an example for round partition with M = 60 bytes, S = 20 bytes, $x$ = 1, and $n$ = 4. 
In this figure, a graph is \textit{first} partitioned into two sub-graphs corresponding to two rounds. For example, V$_{15}$ and V$_{44}$ are partitioned into round 0 and round 1 with their in-coming edges, respectively. 
\textit{Second}, the vID of each vertex (e.g., V$_{15}$) is included into the neighbor lists of its incoming neighbors (e.g., V$_{0}$ and V$_{54}$) to support the scatter-based method. 
As a result, the large neighbor lists of high out-degree vertices (e.g. V${_0}$) are sliced over several rounds, avoiding large packets.

\begin{algorithm}[!t]
 \small
 \label{alg:round_execution}
  \caption{\textbf{Round Execution}}
  \centerline{$\triangleleft$ \textbf{\blackcircledempty{1} Initialization} $\triangleright$}
  load round info and configure round execution; \\
   \centerline{$\triangleleft$ \textbf{\blackcircledempty{2} Load and Send} $\triangleright$}
   \ForEach{vertex $v$ in local}{
     load $v$'s feature vector, network topology, and graph topology; \\
     \If{current node has $v$'s neighbor $u$}
     {
       save a replica in aggregation buffer;   \\
       save \{buffer address, list of all neighbor $u$\} to edge buffer;  \\
     }
     send $v$'s data to remote processing nodes;   \\
  }
  \centerline{$\triangleleft$ \textbf{\blackcircledempty{3} Receive} $\triangleright$}
  receive $v$'s feature vector and graph topology; \\
  save a replica in aggregation buffer; \\
  save \{buffer address, $v$'s neighbors\} to edge buffer; \\ 
    
  \centerline{$\triangleleft$ \textbf{\blackcircledempty{4} Compute} $\triangleright$}
  perform \texttt{aggregate} function using buffer address and neighbors' vID in items of edge buffer;  \\
  perform \texttt{combine} function when aggregation is complete;  \\
  store final combined result to off-chip memory; \\ 

  \centerline{$\triangleleft$ \textbf{\blackcircledempty{5} Synchronization} $\triangleright$}
  synchronize and complete the current round;

\end{algorithm}

Algorithm~\ref{alg:round_execution} demonstrates the round execution method which includes five steps: \blackcircledempty{1} \texttt{Initialization}, where each processing node loads the round info and is configured for a new round.
\blackcircledempty{2} \texttt{Load and Send}, where each processing node loads graph data including feature vectors, graph topology, and network topology, and then sends the graph data to other nodes.
\blackcircledempty{3} \texttt{Receive}, where each processing node receives the replica of feature vector and graph topology from remote nodes into the aggregation buffer and the edge buffer.
\blackcircledempty{4} \texttt{Compute}, where each processing node executes the \texttt{aggregate} function or \texttt{combine} function to process graph data in local or from remote.  
\blackcircledempty{5} \texttt{Synchronization}, where each processing node broadcasts an end signal to others when its workload in the current round is completed, and the current round is terminated after all signals from other nodes are collected. Note that other synchronization mechanisms can also be used for better efficiency. 
Besides, execution overlap technique is utilized in the round execution to improve resource utilization. After step \blackcircledempty{1}, step \blackcircledempty{2}, \blackcircledempty{3}, and \blackcircledempty{4} can be overlapped intra round. Moreover, these three steps can also be overlapped inter round. Furthermore, step \blackcircledempty{4} is able to actively process the graph data locally to keep compute resources busy when no graph data is received.

The round partition and round execution method provide two benefits. \textit{First}, large neighbor lists are sliced, avoiding compute resource underutilization due to the intensive transmission and unacceptable routing latency for a single packet. \textit{Second}, the large volume of replicas are split and processed over a set of rounds, so that replicas in each round can be totally saved in on-chip memory, avoiding the frequent transfer of replicas between on-chip memory and off-chip memory.

\section{Evaluation Methodology} \label{sec:Methodology}

\textbf{Evaluation Tools.} 
We design and implement an in-house simulator to measure execution time in number of cycles. The simulator has a cycle-level model for many microarchitectural components, including multi-bank on-chip buffer, HBM (high bandwidth memory), NVLink, systolic arrays, and so on. 
To measure critical path delay (in cycles) of the router, receive unit, send unit, loader, scheduler, and compute unit, we implement and synthesize these modules in Verilog. We use the Synopsys Design Compiler with the TSMC 12 $nm$ standard VT library for the synthesis and estimate power consumption using Synopsys PrimeTime PX. The slowest module has a critical path delay of 0.83 $ns$ including the setup and hold time, putting MultiGCN comfortably at 1 GHz clock frequency. 
The access latency, energy, and area of the on-chip buffer and FIFO are estimated using Synopsys DesignWare Memory Compiler. 
The access latency and energy of HBM are simulated by Ramulator \cite{Ramulator}, a cycle-accurate DRAM simulator and estimated with 7 pJ/bit as in \cite{HBM_Power}, respectively.
The access latency and energy of NVLink are estimated with around 500 ns as in \cite{NVLink2.0_Latency} and 8 pJ/bit as in \cite{NVLink_Power}, respectively.

\begin{table}[!t]
    \vspace{-5pt}
	\caption{System parameters of MultiGCN @1GHz \& TSMC 12 $nm$.} \label{tab:MultiGCN_spec}
	\vspace{-10pt}
	\centering
	\renewcommand\arraystretch{1.0}
    \resizebox{0.48\textwidth}{!}{
    \begin{tabular}{|c|c|c|c|}
\hline
\multicolumn{4}{|c|}{\textbf{Network Parameters}}\\ \hline
Network Topology  & \#Processing Node & Network Bandwidth & Network Latency \\ \hline
2D Torus & 16     & 600 GB/s  & 500 Cycles          \\ \hline \hline

\multicolumn{4}{|c|}{\textbf{Memory Parameters of Each Processing Node}} \\ \hline
Buffer in Router & Buffer in Send Unit & Buffer in Loader     & Edge Buffer   \\ \hline
1.5 MB           & 512 KB              & 896 KB               & 128 KB             \\ \hline 
Aggregation Buffer & Weight Buffer & Combination Buffer  & HBM Bandwidth \\ \hline
1 MB   & 2 MB      & 256 KB & 256 GB/s \\ \hline \hline

\multicolumn{4}{|c|}{\textbf{Compute Parameters of Each Processing Node}} \\ \hline
\multicolumn{4}{|c|}{8 Reusable Systolic Arrays (each size 1$\times$128)}\\ \hline

\end{tabular}
}
\vspace{-10pt}
\end{table}

\begin{table}[!t]
    %\vspace{-5pt}
	\caption{Graph datasets used in evaluation~\cite{snapnets}.} \label{tab:dataset_spec}
	\vspace{-10pt}
	\centering
	\renewcommand\arraystretch{1.0}
    \resizebox{0.48\textwidth}{!}{
\begin{tabular}{|c|rrrrrrr|}
\hline
\textbf{Name}    & \textbf{$|V|$} & \textbf{$|E|$} & $d_v$ & \textbf{$|h^0|$} & \textbf{$|h^1|$} & \multicolumn{1}{c}{\begin{tabular}[c]{@{}c@{}}\textbf{Topology} \\ \textbf{Size}\end{tabular}} & \multicolumn{1}{c|}{\begin{tabular}[c]{@{}c@{}}\textbf{Feature}\\ \textbf{Size}\end{tabular}} \\ \hline \hline
\multicolumn{8}{|c|}{\begin{tabular}[c]{@{}c@{}} \textbf{Real-world Graphs} \end{tabular}} \\ \hline
Reddit (RD)       & 233K & 114M & 489 & 602 & 128 & 460 MB & 561 MB \\  
Orkut (OR)        & 3M   & 117M & 39  & 500 & 128 & 481 MB & 6 GB   \\  
LiveJournal (LJ)  & 5M   & 69M  & 14 & 500 & 128 & 295 MB & 10 GB  \\   

\hline \hline
\multicolumn{8}{|c|}{\begin{tabular}[c]{@{}c@{}} \textbf{Synthetic Graphs} \end{tabular}} \\ \hline
RMAT-19 (RM19)       & 0.5M  & 16.8M  & 32 & 512 & 128 & 67 MB   & 1 GB \\
RMAT-20 (RM20)       & 1M    & 33.6M  & 32 & 512 & 128 & 134 MB  & 2 GB\\ 
RMAT-21 (RM21)       & 2.1M  & 67.1M  & 32 & 512 & 128 & 269 MB  & 4 GB \\ 
RMAT-22 (RM22)       & 4.2M  & 134M   & 32 & 512 & 128 & 537 MB  & 8 GB \\ 
RMAT-23 (RM23)       & 8.4M  & 268M   & 32 & 512 & 128 & 1074 MB & 16 GB \\

\hline
\end{tabular}
    }
    \vspace{-8pt}
\end{table}

\begin{figure*}[!hptb] 
    \centering
    \includegraphics[page=1, width=0.88\textwidth]{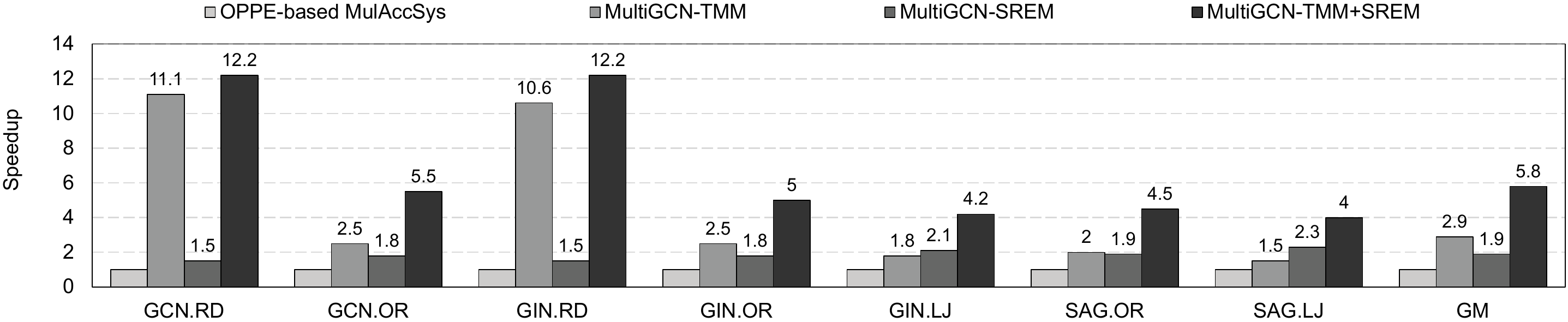}
    \vspace{-11pt}
    \caption{Performance comparison between OPPE-based MulAccSys and MultiGCN (normalized to OPPE-based MulAccSys).}
    \label{fig:speedup}
    \vspace{-10pt}
\end{figure*}

\begin{table*}[!htpb]
    \centering
    \caption{Utilization ratio (\%) of network bandwidth, DRAM bandwidth, and compute unit of OPPE-based MulAccSys and MultiGCN.} \label{tab:speedup_analysis}
    \vspace{-10pt}
    
    %%%%%%%%%%%%%%%%%%%%%%%%%%%%%%%%%%%%%%%%%
    \begin{minipage}{0.88\textwidth}
    \centering
    \setlength\tabcolsep{2pt}
	\centering
	\renewcommand\arraystretch{1.0}
    \resizebox{0.88\textwidth}{!}{
\begin{tabular}{|c|CCC|CCC|CCC|CCC|}
\hline
       & \multicolumn{3}{c|}{\textbf{OPPE-based MulAccSys}} 
       & \multicolumn{3}{c|}{\textbf{MultiGCN-TMM}}                 
       & \multicolumn{3}{c|}{\textbf{MultiGCN-SREM}}        
       & \multicolumn{3}{c|}{\textbf{MultiGCN-TMM+SREM}}     
       \\ \hline \hline
       & \multicolumn{1}{c} {\makecell{Network\\ Bandwidth}}
       & \multicolumn{1}{c} {\makecell{DRAM\\ Bandwidth}} 
       & \multicolumn{1}{c|}{\makecell{Compute\\ Unit}}
       & \multicolumn{1}{c} {\makecell{Network\\ Bandwidth}}
       & \multicolumn{1}{c} {\makecell{DRAM\\ Bandwidth}}
       & \multicolumn{1}{c|}{\makecell{Compute\\ Unit}}
       & \multicolumn{1}{c} {\makecell{Network\\ Bandwidth}}
       & \multicolumn{1}{c} {\makecell{DRAM\\ Bandwidth}}
       & \multicolumn{1}{c|}{\makecell{Compute\\ Unit}}
       & \multicolumn{1}{c} {\makecell{Network\\ Bandwidth}}
       & \multicolumn{1}{c} {\makecell{DRAM\\ Bandwidth}}
       & \multicolumn{1}{c|}{\makecell{Compute\\ Unit}}
       \\ \hline

GCN.RD	& 19	& 12	& 2	& 4	& 29	& 20	& 28	& 17	& 3	& 87	& 14	& 22	\\ \hline
GCN.OR	& 17	& 15	& 6	& 6	& 41	& 16	& 31	& 20	& 12	& 69	& 32	& 35	\\ \hline
GCN.LJ	& 16	& 20	& 14	& 7	& 41	& 26	& 34	& 23	& 30	& 60	& 31	& 68	\\ \hline
GIN.RD	& 19	& 12	& 2	& 4	& 21	& 21	& 28	& 17	& 3	& 87	& 14	& 24	\\ \hline
GIN.OR	& 17	& 15	& 8	& 6	& 41	& 21	& 31	& 19	& 15	& 62	& 29	& 41	\\ \hline
GIN.LJ	& 15	& 19	& 19	& 7	& 41	& 34	& 33	& 22	& 40	& 51	& 26	& 80	\\ \hline
SAG.RD	& 19	& 19	& 9	& 8	& 39	& 17	& 45	& 29	& 20	& 76	& 36	& 39	\\ \hline
SAG.OR	& 16	& 18	& 10	& 7	& 42	& 21	& 31	& 21	& 20	& 60	& 31	& 46	\\ \hline
SAG.LJ	& 15	& 23	& 21	& 8	& 40	& 32	& 34	& 24	& 47	& 53	& 30	& 84	\\ \hline
GM	& 17	& 17	& 8	& 6	& 37	& 22	& 33	& 21	& 15	& 66	& 26	& 44	\\ \hline

    \end{tabular}
    }
    \end{minipage}\hfill
%%%%%%%%%%%%%%%%%%%%%%%%%%%%%%%%%%%%%%%%%
    \begin{minipage}{0.1\textwidth}
    \centering
    \includegraphics[page=1, width=0.5\textwidth]{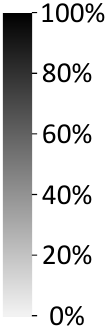}
    \end{minipage}\hfill
    \vspace{-10pt}
\end{table*}

\textbf{Baselines and System Configurations.} 
To demonstrate the advantages of MultiGCN, we compare MultiGCN with a single-node GCN accelerator (i.e., AWB-GCN~\cite{AWB-GCN}) using identical hardware resources, two GPU-based solutions (i.e., PyG~\cite{PyTorch_Geometric} and GNNAdvisor\cite{GNNAdvisor}) running on T4 GPU, OPPE-based MulAccSys, and OPPR-based MulAccSys. 
Three configurations of MultiGCN are evaluated to assess MultiGCN. 
The first configuration is MultiGCN only employing the topology-aware multicast mechanism (TMM), denoted by \textbf{MultiGCN-TMM}. 
The second configuration is MultiGCN only employing the scatter-based round execution mechanism (SREM), denoted by \textbf{MultiGCN-SREM}.
The last configuration is MultiGCN employing both TMM and SREM, denoted by \textbf{MultiGCN-TMM+SREM}. 
All these configurations use the system parameters described in Table~\ref{tab:MultiGCN_spec}. 

\textbf{Workloads.}
We implement three well-known GCNs in MultiGCN, namely GCN~\cite{1stChebNet}, GINConv (GIN)~\cite{GINConv}, and GraphSAGE (SAG)~\cite{GraphSage}. Due to the long simulation time on large-scale graphs, we simulate only the first layer of these models. Since runtime characteristics of GCNs are input-dependent, we use several real-world and synthetic graphs as inputs to each GCN model, as shown in Table~\ref{tab:dataset_spec}. 
Topology size of graph refers to the total size of edges, $|E|*4$ Bytes.
Feature size of graph refers to the total size of feature vectors, $|V|*|h^0|*4$ Bytes.

\section{Results}\label{overall_result}

\subsection{Overall Results} \label{result:performance}

\textbf{Performance.} Figure~\ref{fig:speedup} compares the performance of the proposed MultiGCN against that of OPPE-based MulAccSys.  
In this figure, the last set of bars, labeled GM, indicates the geometric mean across all workloads.
Our evaluation shows that MultiGCN with only the TMM mechanism or only SREM mechanism outperforms OPPE-based MulAccSys by 2.9$\times$ or 1.9$\times$ on average. 
When both mechanisms are employed, MultiGCN achieves 4$\sim$12$\times$ speedup over OPPE-based MulAccSys, and 5.8$\times$ on average.

To provide more insights into the performance improvement of MultiGCN, Table~\ref{tab:speedup_analysis} shows the utilization ratios of network bandwidth, DRAM bandwidth, and compute unit in MultiGCN.
Compared with OPPE-based MulAccSys, the utilization ratio of network bandwidth, DRAM bandwidth and compute unit of MultiGCN-TMM+SREM improve by 3.88$\times$, 1.53$\times$, and 7.33$\times$ on average, respectively. This points to the main contributor of the large speedup achieved by our design: the TMM mechanism and the SREM mechanism.

\begin{table}[!t]
    \vspace{-5pt}
	\caption{Characteristics of processing node @1GHz \& TSMC 12 $nm$.} \label{tab:chip_area_power}
	%\vspace{-10pt}
	\centering
	\renewcommand\arraystretch{1.0}
    \resizebox{0.46\textwidth}{!}{
\begin{tabular}{|l|l|r|r|r|r|}
\hline
\multicolumn{2}{|l|}{\begin{tabular}[c]{@{}c@{}}\textbf{Component or Block}\end{tabular}} & \begin{tabular}[r]{@{}c@{}}\textbf{Area ($mm^2$)}\end{tabular} & \textbf{\%} & \begin{tabular}[r]{@{}c@{}}\textbf{Power ($mW$)}\end{tabular}  & \textbf{\%} \\ \hline \hline
\multicolumn{2}{|l|}{{Processing Node}}    &12.4	&100	&3671.13	&100    \\ \hline \hline
\multicolumn{6}{|c|}{\begin{tabular}[c]{@{}c@{}} \textbf{Breakdown by Functional Block} \end{tabular}} \\ \hline
\multicolumn{2}{|l|}{Edge Buffer}     &0.23	&1.88	&9.03	&0.25    \\
\multicolumn{2}{|l|}{Aggregation Buffer}   &1.87	&15.06	&578.3	&15.75    \\
\multicolumn{2}{|l|}{Weight Buffer}   &3.74	&30.11	&614.13	&16.73     \\
\multicolumn{2}{|l|}{Combination Buffer}   &0.47	&3.76	&551.42	&15.02    \\
\multicolumn{2}{|l|}{Compute Unit}    &0.84	&6.8	&650.63	&17.72   \\
\multicolumn{2}{|l|}{Router}          &2.8	&22.59	&689.45	&18.78    \\
\multicolumn{2}{|l|}{Loader}          &1.52	&12.24	&320.51	&8.73    \\
\multicolumn{2}{|l|}{Send Unit}          &0.93	&7.53	&257.6	&7.02   \\
\multicolumn{2}{|l|}{Scheduler}       &4.73E-04	&0.00 &	0.00 &	0.00   \\
\multicolumn{2}{|l|}{Others}          &1.89E-03	&0.02 &	0.04 & 0.00    \\  \hline
\end{tabular}
    }

\end{table}

\textbf{Area and Power.} 
Table~\ref{tab:chip_area_power} provides the detailed characteristics of MultiGCN. 
The area and power of each processing node are 12.4 $mm^2$ and 3671.13 $mW$ respectively.
The buffers including edge buffer, aggregation buffer, weight buffer, and combination buffer occupy most area of the processing node and accounts for 48\% power of the processing node. 
The area and power produced by the compute unit are 6.8\% and 17.72\%. For the computation precision, we use 32-bit fixed point which is enough for accurate GCN inference. 
The area and power produced by router are 22.59\% and 18.78\% due to the large routing buffer and heavy packet transmissions.  

\begin{figure}[!t] 
    \vspace{-5pt}
    \centering
    \includegraphics[page=1, width=0.5\textwidth]{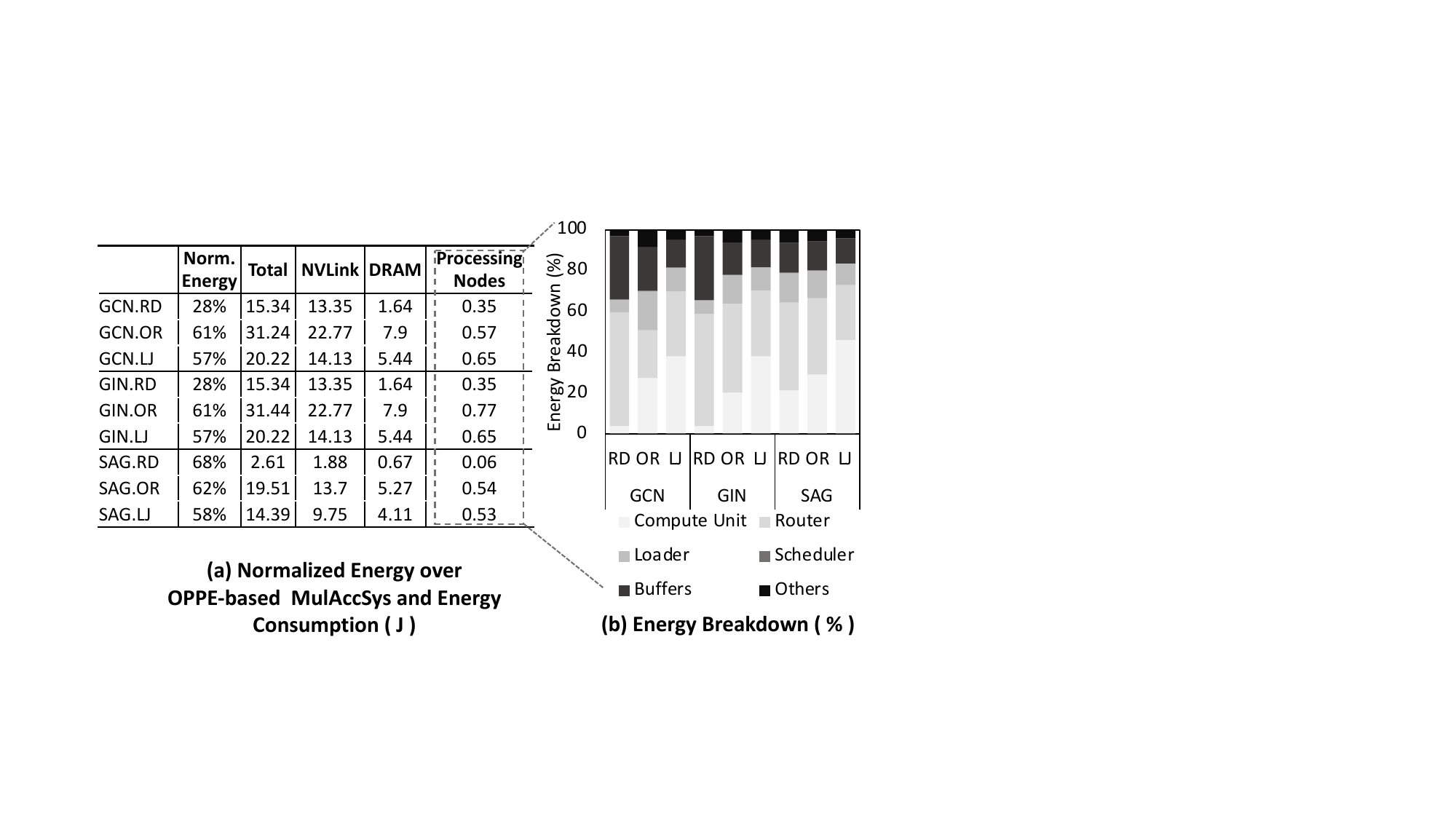}
    \vspace{-15pt}
    \caption{Energy: (a) Normalized energy over OPPE-based MulAccSys and energy details; (b) Energy breakdown of processing nodes.}
    \label{fig:energy}
\end{figure}

\textbf{Energy and its Breakdown.} Figure~\ref{fig:energy} shows the energy consumption of MultiGCN-TMM+SREM in detail. Figure~\ref{fig:energy}(a) depicts that MultiGCN costs only 28\%$\sim$68\% energy of OPPE-based MulAccSys. The energy consumed by network (i.e., NVLink) is larger than DRAM and processing nodes in all cases, since the feature vector loaded from DRAM can be shared across multiple remote processing nodes. 
Figure~\ref{fig:energy}(b) shows that the compute unit, router and buffers consume most energy of the processing nodes.

\subsection{Effect and Overhead of Optimizations} \label{resulst:optimization_analysis}
To dissect the effect of our optimizations, the normalized network transmission and DRAM access of MultiGCN (normalized to OPPE-based MulAccSys) are shown in Table~\ref{tab:reduction}. The reduction of redundant transmissions and DRAM
accesses as well as the overhead analysis of these optimizations are also shown in Table~\ref{tab:overhead}.

\begin{table}[!t]
	\caption{Normalized network transmission and DRAM access of MultiGCN (normalized to OPPE-based MulAccSys).} \label{tab:reduction}
	\vspace{-10pt}
	\centering
	\renewcommand\arraystretch{1.0}
    \resizebox{0.48\textwidth}{!}{
\begin{tabular}{|c|c|c|c|c|c|c|}
\hline
       & \multicolumn{2}{c|}{\textbf{MultiGCN-TMM}} & \multicolumn{2}{c|}{\begin{tabular}[c]{@{}c@{}}\textbf{MultiGCN-SREM}\end{tabular}} & \multicolumn{2}{c|}{\begin{tabular}[c]{@{}c@{}}\textbf{MultiGCN-SREM}\\\textbf{+TMM}\end{tabular}} \\ \hline \hline
       & Trans.     & Access     & Trans.       & Access       & Trans.       & Access     \\ \hline
GCN.RD	&2\%	&21\%	&100\%	&93\%	&37\%	&10\%	\\
GCN.OR	&14\%	&112\%	&100\%	&72\%	&75\%	&39\%	\\
GCN.LJ	&25\%	&114\%	&100\%	&53\%	&79\%	&33\%	\\
GIN.RD	&2\%	&16\%	&100\%	&93\%	&37\%	&10\%	\\
GIN.OR	&14\%	&112\%	&100\%	&72\%	&75\%	&39\%	\\
GIN.LJ	&25\%	&118\%	&100\%	&53\%	&79\%	&33\%	\\
SAG.RD	&20\%	&102\%	&100\%	&63\%	&88\%	&41\%	\\
SAG.OR	&22\%	&116\%	&100\%	&60\%	&81\%	&38\%	\\
SAG.LJ	&35\%	&115\%	&100\%	&48\%	&86\%	&33\%	\\ \hline
GM	&13\%	&75\%	&100\%	&66\%	&68\%	&27\%	\\
\hline

\end{tabular}
    }
\vspace{-10pt}
\end{table}

\textbf{Effect.}
The TMM mechanism helps eliminate the redundant transmissions. 
Table~\ref{tab:reduction} shows that the network transmission of MultiGCN-TMM is only 13\% that of OPPE-based MulAccSys. 
This is because a single packet containing a replica of the feature vector is sent to many other processing nodes that also request it via multicast. 
Note that the number of DRAM access in RD dataset decreases, but extra DRAM accesses are introduced in most datasets.
This is because a feature vector in RD dataset loaded from DRAM can be shared by many remote processing nodes due to its extremely high average degree (e.g., 489). 
The SREM mechanism avoids the frequent transfer of replicas between on-chip memory and off-chip memory. Specifically, the rounds are properly partitioned so that the replicas of all vertices and intermediate results in a round always stays on-chip until the computation is done.
Compared with OPPE-based MulAccSys, MultiGCN-SREM introduces only 66\% number of DRAM accesses on average. 

Table~\ref{tab:reduction} shows that when these two mechanisms are employed, both the network transmission and number of DRAM accesses are reduced significantly to only 68\% and 27\% on average, respectively. 
Note that the effect of TMM mechanism is hurt by the SREM mechanism because each round may introduce a multicast of the same feature vector. In contrast, TMM mechanism promotes the effect of SREM mechanism since a feature vector loaded from DRAM can be multicast to and shared by many remote processing nodes.
Table~\ref{tab:overhead} depicts that MultiGCN-TMM+SREM reduces 32\% redundant network transmissions and 100\% of redundant DRAM accesses on average compared to OPPE-based MulAccSys.

\textbf{Overhead.} The main optimization overheads are the extras of transmission latency and preprocesing time for round partition, but all of them are small, only 0.21\% and 6.1\% on average, as shown in the last two columns in Table~\ref{tab:overhead}. 
Note that as more redundant transmissions are reduced, the network topology and graph topology information in the packet increases transmission latency.
The round partition accounts for less than 12\% time of the graph mapping because it can be coupled into the process of graph mapping. Besides, it is a one-time overhead for each dataset that can be amortized over the execution of different GCN models.

\begin{table}[!t]
    \caption{Reduction of redundant transmission and redundant DRAM access, extra transmission latency, and extra preprocessing time compared with OPPE-based MulAccSys.}
	\vspace{-10pt}
	\label{tab:overhead}
	\centering
	\renewcommand\arraystretch{1.0}
    \resizebox{0.49\textwidth}{!}{

\begin{tabular}{|c|c|c|c|c|}
\hline
  & \multicolumn{4}{c|}{\textbf{MultiGCN-TMM+SREM}} \\ \hline \hline
  & \multicolumn{1}{c|}{\begin{tabular}[c]{@{}c@{}}Redundant \\ Transmission\\\end{tabular}} 
  & \multicolumn{1}{c|}{\begin{tabular}[c]{@{}c@{}}Redundant \\ DRAM access\\\end{tabular}} 
  & \multicolumn{1}{c|}{\begin{tabular}[c]{@{}c@{}}Transmission \\Latency\end{tabular}} 
  & \multicolumn{1}{c|}{\begin{tabular}[c]{@{}c@{}}Round Partition \\Time\end{tabular}} 
  \\ \hline 
GCN.RD	&-64\%	&-100\%	&+0.52\%	&+6.6\%		\\
GCN.OR	&-30\%	&-100\%	&+0.15\%	&+12\%		\\
GCN.LJ	&-30\%	&-100\%	&+0.13\%	&+2.8\%		\\
GIN.RD	&-64\%	&-100\%	&+0.52\%	&+6.6\%		\\
GIN.OR	&-30\%	&-100\%	&+0.15\%	&+12\%		\\
GIN.LJ	&-30\%	&-100\%	&+0.13\%	&+2.8\%		\\
SAG.RD	&-17\%	&-100\%	&+0.07\%	&+6.6\%		\\
SAG.OR	&-25\%	&-100\%	&+0.11\%	&+12\%		\\
SAG.LJ	&-24\%	&-100\%	&+0.1\%	&+2.8\%		\\ \hline
GM	&-32\%	&-100\%	&+0.21\%	&+6.1\%		\\

\hline
\end{tabular}
    }
\vspace{-5pt}
\end{table}

\subsection{Comparisons with the State of the Arts} \label{sec:result_comparison}

The performance of MultiGCN (1 node) is slightly lower than that of GNNAdvisor running on one T4 GPU, 0.7$\times$ on average.
However, MultiGCN aims to scale single-node accelerator to accelerate GCNs on large-scale graphs efficiently, such as average 3.4$\times$ speedup of MultiGCN (4 nodes) over GNNAdvisor.

\begin{figure}[!t]
    \centering
    \includegraphics[page=1, width=0.46\textwidth]{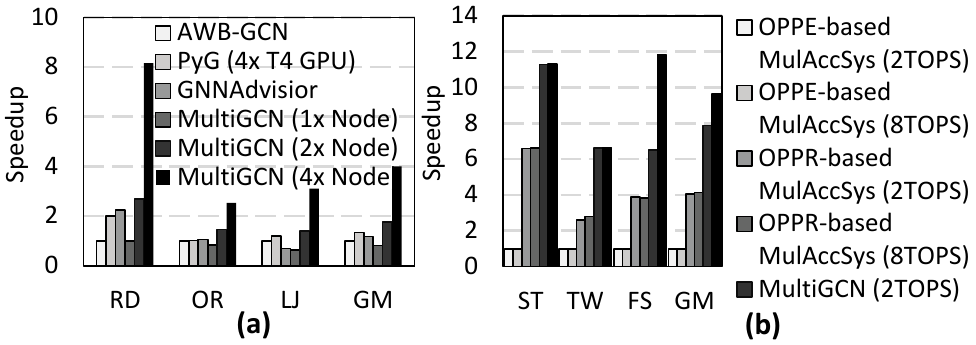}
    \vspace{-10pt}
    \caption{Comparisons with the state-of-the-arts: (a) Speedup of PyG (4 T4 GPUs), GNNAdvisor (1 T4 GPU), MultiGCN (2 nodes), and MultiGCN (4 nodes) over AWB-GCN; (b) Speedup of MultiGCN (128 nodes) over OPPE-based and OPPR-based MulAccSyses (128 nodes).}
    \label{fig:comparison}
\end{figure}

To demonstrate the advantages of MultiGCN, we compare MultiGCN against the state-of-the-arts.
Figure~\ref{fig:comparison}(a) shows that the average speedup of MultiGCN (1 node) is slightly lower than that of AWB-GCN and GNNAdvisor.
However, MultiGCN aims to scale single-node accelerator to accelerate GCNs on large-scale graphs efficiently, such as on average 4$\times$ and 3.4$\times$ speedup of MultiGCN (4 nodes) over AWB-GCN and GNNAdvisor respectively. Moreover, the peak performance, DRAM bandwidth, and maximum power of MultiGCN (1 node) are respectively only 25\%, 85\%, and 5\% of T4 GPU.
MultiGCN (4 nodes) achieves 4$\times$ speedup on average over PyG with 4 T4 GPUs. This is because GPUs aim at workloads with regular execution pattern, but they cannot efficiently tackle irregular execution patterns of GCNs~\cite{HyGCN,AWB-GCN,EnGN,GraphDynS,Graphicionado}. 
Note that GPU performance of PyG is estimated by accumulating the kernel execution time which does not account for the memory copy time and system stack overhead. 
Besides, PyG leverages mini-batch to make each GPU execute inference independently, so that inter-GPU communications are eliminated. 
However, mini-batching multiplies data volume in system due to massive copies of neighboring feature vectors in each GPU. 
Figure~\ref{fig:comparison}(b) shows that MultiGCN (128 nodes and 8 TOPS) achieves average 9.6$\times$ and 2.3$\times$ speedup over OPPE-based MulAccSys (128 nodes and 8 TOPS) and OPPR-based MulAccSys (128 nodes and 8 TOPS) respectively due to less network transmissions and DRAM accesses.
MultiGCN has different speedup on the FS dataset when compute capability increases because the number of network transmissions of FS is little while the number of compute for local data processing is large.

\begin{figure}[!t]
    \centering
    \includegraphics[page=1, width=0.46\textwidth]{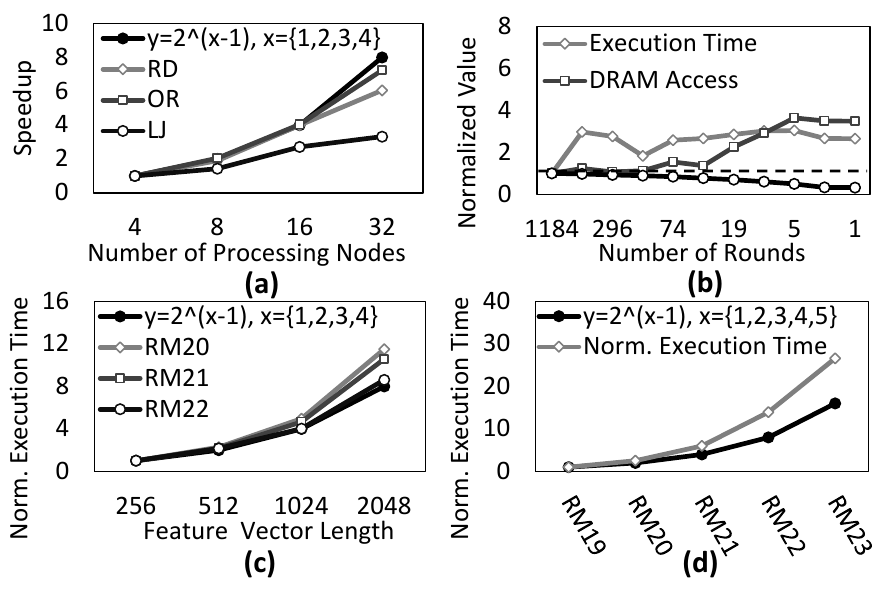}
    \vspace{-10pt}
    \caption{Hardware sensitivity: (a) Speedup over number of processing nodes; (b) Normalized execution time, normalized amount of network transmission, and normalized amount of DRAM access across number of rounds.
    Graph characteristics sensitivity: Normalized execution time across (c) feature vector length and (d) scale of vertex number. }
    \label{fig:exploration}
\end{figure}

\subsection{Exploration of Design}

We conduct several experiments on the GCN model to explore our architecture design in terms of hardware sensitivity and graph characteristic sensitivity as follows.
Due to the long simulation time, we simulate the processing for 10\% of vertices for each experiment in this subsection.

\textbf{Hardware Sensitivity.} 
\uline{\textit{First}}, Figure~\ref{fig:exploration}(a) illustrates the speedup across different number of nodes of the 2D torus network topology. 
MultiGCN gains performance improvement on RD and OR datasets as the number of nodes increases, and the speedup remains linear as the number of nodes increases to 32. Limited by the network bandwidth, the performance gain on LJ dataset gradually decreases as the number of node increases. This is due to the low reusability of feature vector in the network transmission since the average vertex degree of the LJ dataset is low.
\uline{\textit{Second}}, Figure~\ref{fig:exploration}(b) shows that the amount of network transmission decreases as the number of round decreases in the processing of LJ dataset because the number of multicast for the same feature vector decreases. Besides, the execution time and number of DRAM accesses are variable across different numbers of rounds. 
This leaves room for further optimizing the design to fit different requirements, which will be explored in our future work.

\textbf{Graph Characteristic Sensitivity.} 
\uline{\textit{First}}, the length of feature vector (i.e., $|h^0|$) doubles in Figure~\ref{fig:exploration}(c), which means the amount of workload in both the \texttt{Aggregation} phase and \texttt{Combination} phase double, and network transmission increases even more. However, the execution time increases to more than 2$\times$. In particular, the performance is sensitive to the length of the feature vector, which is mainly because the increased network traffic imposes heavier burdens on network. 
\uline{\textit{Second}}, the number of vertex in graph doubles in Figure~\ref{fig:exploration}(d), which means the amount of workload in both the \texttt{Aggregation} phase and \texttt{Combination} phase and network transmission doubles too. However, the execution time increase to more than 2$\times$. That is, the performance is sensitive to the scale of vertex number, since a larger scale of vertex number with the same average degree means higher graph sparsity, which hinders the performance.

\section{Related Work} \label{sec:related_work}

Many software frameworks for GCNs have been developed to relieve programming efforts while improving performance on modern architectures~\cite{PyTorch_Geometric,DGL,AliGraph,Dorylus,P3}. 
For example, PyG~\cite{PyTorch_Geometric} is built upon PyTorch for easy implementation of graph neural networks (GNNs). It consists of easy-to-use mini-batch loaders for giant graphs and multi-GPU support. 
GNNAdvisor~\cite{GNNAdvisor} propose an adaptive and efficient runtime system for GNN acceleration on GPUs.
DGL~\cite{DGL} is a scalable GNN framework that simplifies the development of efficient GNN-based training and inference programs at a large scale in multi-GPU and distributed systems. 
Unfortunately, the distinct execution pattern in GCNs causes processing inefficiencies on conventional architectures. 
Therefore, GCNs demand specialized architecture design.

Characterizing the execution pattern and execution semantic of GNNs on GPUs is important for both software and hardware optimizations for GNNs, which has been extensively studied in previous work~\cite{GCN_Characterization_CAL,GNN_Computational_Graph,GNN_Architectural_Implication,Characterizing_HGNNs,character_Distributed_GNN,Understanding_GNN}.
Yan et al. \cite{GCN_Characterization_CAL} reveal the computation and memory accessing pattern of GCNs on GPU. Yan et al. \cite{Characterizing_HGNNs} disclose the execution pattern and execution semantic of heterogeneous graph neural networks (HGNNs) on GPU.
Zhang et al. \cite{GNN_Computational_Graph} focus on understanding the computational graph of GNN, from the perspective of computation, IO and memory. 
Zhang et al. \cite{GNN_Architectural_Implication} characterize the computation of a large portion of GNN variants concerning general-purpose and application-specific architectures.

Hardware acceleration for GCNs has been recently explored. Many single-node domain-specific architectures and frameworks have been designed for GCN acceleration~\cite{HyGCN,AWB-GCN,GCNAX,Rubik,BoostGCN,GNNerator,GNNIE,BlockGNN,EnGN,I-GCN,gnn_hardware_accelerator,GraphACT,asap_GCN_accelerator}. For example, HyGCN \cite{HyGCN} proposes a hybrid architecture to address the hybrid execution pattern of GCNs. 
AWB-GCN \cite{AWB-GCN} targets workload imbalance in the acceleration of GCNs. 
GCNAX~\cite{GCNAX} proposes a flexible and optimized dataflow for GCNs that simultaneously improves resource utilization and reduces data transfer.
RE-FLIP \cite{RE-FLIP} designs PIM-featured crossbar architectures to build a unified architecture to perform the hybrid execution pattern of GCNs.

The ever-growing scale of graphs has posed new challenges that single-node accelerators cannot sufficiently address. Thus, a multi-node acceleration system is highly desirable.
Although a straightforward multi-node design for large-scale GCNs follows Tesseract~\cite{Tesseract} or other Tesseract-based architectures\cite{GraphP,GraphQ}, it suffers from two inefficiencies including a vast of redundant transmissions and off-chip memory accesses.
To this end, we propose MultiGCN, an efficient MultiAccSys for large-scale GCNs.

\section{Conclusion} \label{sec:conclusion}

In this work, we aim to scale the single-node GCN accelerator to accelerate execution of GCNs on large-scale graphs.
We first characterize the communication pattern and challenges of multi-node acceleration for GCNs.
Guided by our observations, we then propose MultiGCN, an efficient MultiAccSys for large-scale GCNs that trades network latency for network bandwidth.
MultiGCN achieves 4$\sim$12$\times$ and 2.5$\sim$8$\times$ speedup over baseline MultiAccSys and multi-GPU solution respectively.
Designing multi-node acceleration systems is vital to enable practical execution of GNNs on real-world large-scale graphs. We believe our work will draw more attention to the design of domain-specific processor clusters for increasingly important GNNs and graph-structured data.

\ifCLASSOPTIONcompsoc
  \section*{Acknowledgments}
\else
  \section*{Acknowledgment}
\fi
We sincerely thank Prof. Guang R. Gao for his guidance and contribution to this work.
We also would like to express our gratitude to all reviewers' constructive comments for helping us polish this paper.
This work was supported by the National Natural Science Foundation of China (Grant No. 61732018, and 61872335), Austrian-Chinese Cooperative R\&D Project (FFG and CAS) (Grant No. 171111KYSB20200002), CAS Project for Young Scientists in Basic Research (Grant No. YSBR-029), Open Research Projects of Zhejiang Lab (NO. 2022PB0AB01), and CAS Project for Youth Innovation Promotion Association.

\ifCLASSOPTIONcaptionsoff
  \newpage
\fi

%%%%%%%%% -- BIB STYLE AND FILE -- %%%%%%%%
\bibliographystyle{IEEEtran}
%\balance
\bibliography{refs}
%%%%%%%%%%%%%%%%%%%%%%%%%%%%%%%%%%%%

\vspace{-40pt}
\begin{IEEEbiography}[{\includegraphics[width=1in, height=1.25in, clip, keepaspectratio]{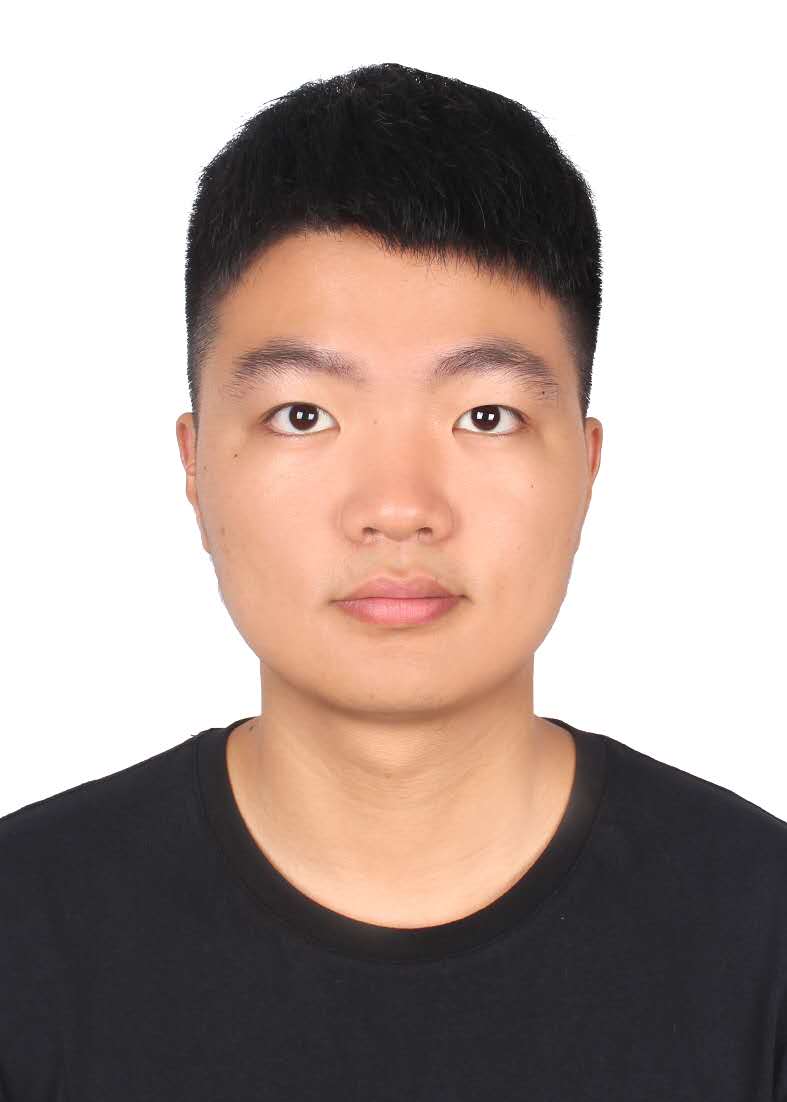}}] {Gongjian Sun} received the B.S. degree from University of Chinese Academy of Sciences, Beijing, China in 2019. He is currently a postgraduate at Institute of Computing Technology, Chinese Academy of Sciences, Beijing, China. His current research interests include high-throughput computer architecture and graph-based hardware accelerator.
\end{IEEEbiography}
%\vspace{-30pt}
\begin{IEEEbiography}[{\includegraphics[width=1in, height=1.25in, clip, keepaspectratio]{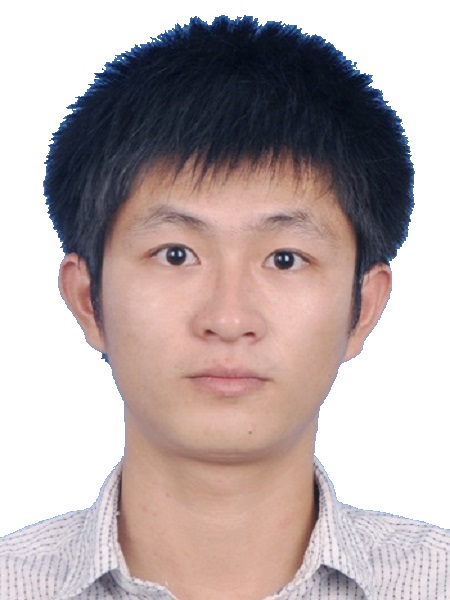}}] {Mingyu Yan} received his Ph.D. degree from University of Chinese Academy of Sciences, Beijing, China in 2020. He is currently an Assistant Professor at Institute of Computing Technology, Chinese Academy of Sciences, Beijing, China. His current research interests include graph-based hardware accelerator and high-throughput computer architecture.
\end{IEEEbiography}
\vspace{-30pt}
\begin{IEEEbiography}[{\includegraphics[width=1in, height=1.25in, clip, keepaspectratio]{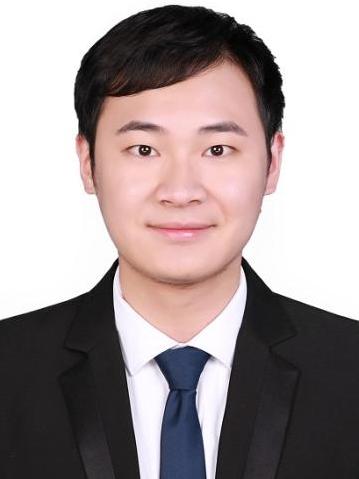}}] {Duo Wang} received his B.S. degree from Southeast University, Nanjing, China in 2018. He is currently a Ph.D. candidate at Institute of Computing Technology, Chinese Academy of Sciences, Beijing, China. His current research interests include high-performance computer architecture and software simulation.
\end{IEEEbiography}
\vspace{-30pt}
\begin{IEEEbiography}[{\includegraphics[width=1in, height=1.25in, clip, keepaspectratio]{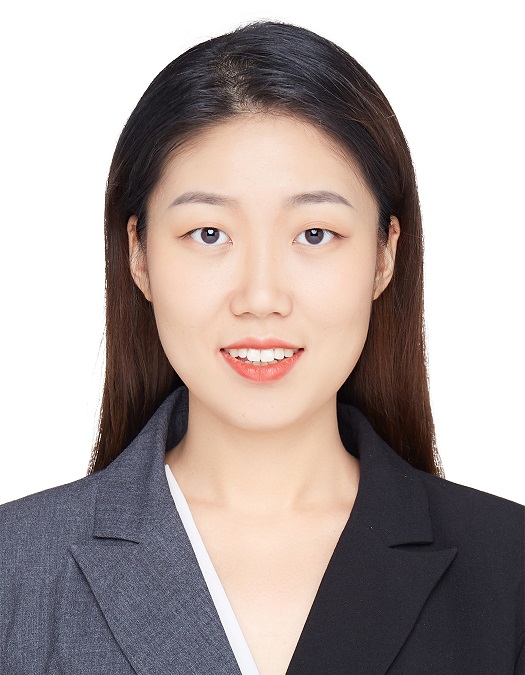}}] {Han Li} received the B.S. degree from Jilin University, Changchun, China in 2016. She is currently a Ph.D. candidate at Institute of Computing Technology, Chinese Academy of Sciences, Beijing, China. Her current research interests include computer architecture and graph-based hardware accelerator.
\end{IEEEbiography}
\vspace{-30pt}
\begin{IEEEbiography}[{\includegraphics[width=1in, height=1.25in, clip, keepaspectratio]{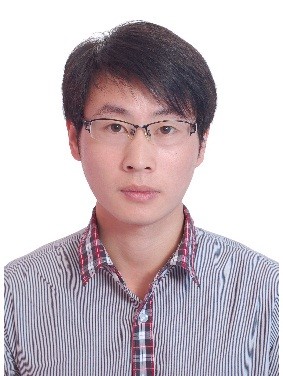}}] {Wenming Li} received the Ph.D. degree in computer architecture from Institute of Computing Technology, Chinese Academy of Sciences, Beijing, in 2016. He is currently an associate professor in Institute of Computing Technology, Chinese Academy of Sciences, Beijing. His main research interests include high-throughput processor architecture, dataflow architecture and software simulation. 
\end{IEEEbiography}
\vspace{-35pt}
\begin{IEEEbiography}[{\includegraphics[width=1in, height=1.25in, clip, keepaspectratio]{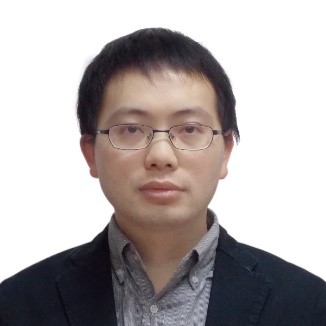}}] {Xiaochun Ye} received his Ph.D. degree in computer architecture from Institute of Computing Technology, Chinese Academy of Sciences, Beijing, in 2010. He is currently an associate professor in Institute of Computing Technology, Chinese Academy of Sciences, Beijing. His main research interests include high-performance computer architecture and software simulation.
\end{IEEEbiography}
\vspace{-40pt}
\begin{IEEEbiography}[{\includegraphics[width=1in, height=1.25in, clip, keepaspectratio]{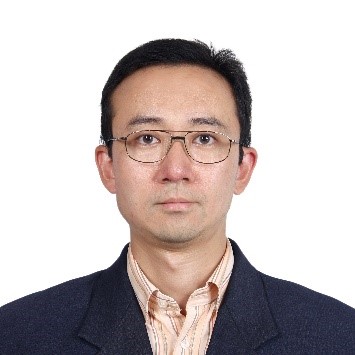}}] {Dongrui Fan} received his Ph.D. degree in computer architecture from Institute of Computing Technology, Chinese Academy of Sciences, Beijing, in 2005. He is currently a professor and Ph.D. supervisor in Institute of Computing Technology, Chinese Academy of Sciences, Beijing. His main research interests include high-throughput computer architecture and high-performance computer architecture. 
\end{IEEEbiography}
\vspace{-40pt}
\begin{IEEEbiography}[{\includegraphics[width=1in,height=1.25in,clip,keepaspectratio]{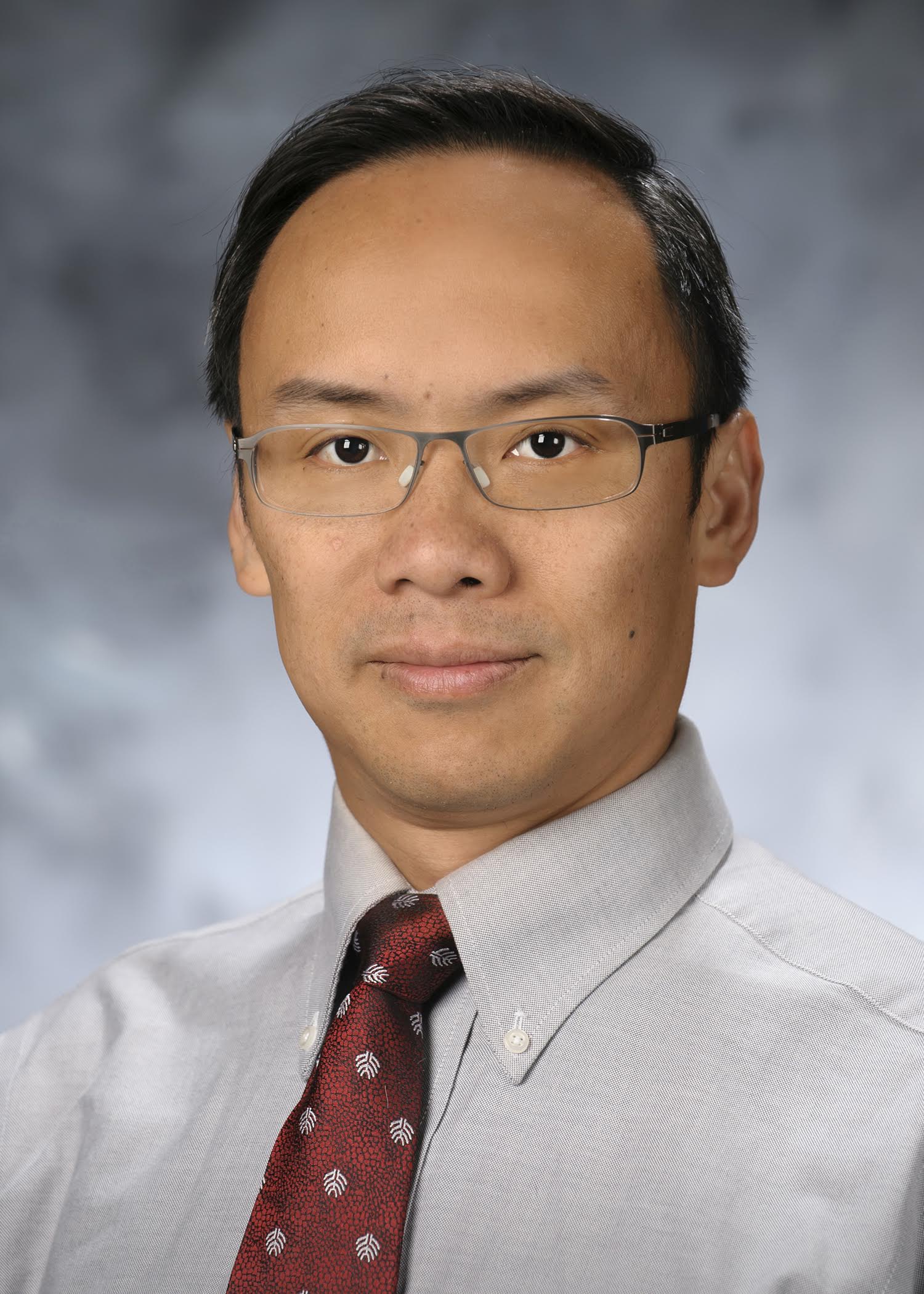}}]{Yuan Xie} received his Ph.D. degrees from Electrical Engineering Department, Princeton University, Princeton, NJ, USA in 2002. He was a Professor with Pennsylvania State University, State College, PA, USA, from 2003 to 2014. He is currently a Professor with the Department of Electrical and Computer Engineering, University of California at Santa Barbara, Santa Barbara, CA, USA. 
\end{IEEEbiography}
%

%\input{tex/ResponseV1}

\end{document}